\documentclass[noshowpacs,prl,twocolumn,english]{revtex4}

\usepackage{graphicx}
\usepackage{dcolumn}
\usepackage{bm}
\usepackage{natbib}
\usepackage{amsmath}
\usepackage{appendix}
\usepackage{here}
\usepackage{color}
\usepackage{ulem}
\usepackage{epstopdf}
\usepackage{pdfpages}
\usepackage[a4paper, total={7in, 9.5in}]{geometry}

\makeatletter

 \@ifundefined{textcolor}{}
 {%
   \definecolor{BLACK}{gray}{0}
   \definecolor{WHITE}{gray}{1}
   \definecolor{RED}{rgb}{1,0,0}
   \definecolor{GREEN}{rgb}{0,1,0}
   \definecolor{BLUE}{rgb}{0,0,1}
   \definecolor{CYAN}{cmyk}{1,0,0,0}
   \definecolor{MAGENTA}{cmyk}{0,1,0,0}
   \definecolor{YELLOW}{cmyk}{0,0,1,0}
 }

\makeatother

\usepackage{babel}
\begin{document}

\title{Quantum Optimization of Fully-Connected Spin Glasses}

\author{Davide Venturelli$^{1,2\ast}$, Salvatore Mandr\`a$^{3}$, Sergey Knysh$^{1,4}$, Bryan O'Gorman$^{1}$, Rupak Biswas$^{1}$ and Vadim Smelyanskiy$^{1\ast}$}
\affiliation{$^{1}$NASA Ames Research Center Quantum Artificial Intelligence Laboratory (QuAIL), Mail Stop 269-1, 94035 Moffett Field CA}
\affiliation{$^{2}$USRA Research Institute for Advanced Computer Science (RIACS), 615 National, 94043 Mountain View CA}
\affiliation{$^{3}$Harvard University, Department of Chemistry and Chemical Biology, 12 Oxford Street, 02139 Cambridge MA}
\affiliation{$^{4}$Stinger Ghaffarian Technologies Inc., 7701 Greenbelt Rd., Suite 400, Greenbelt, MD 20770}
\affiliation{$^\ast$Correspondence should be addressed to:  davide.venturelli@nasa.gov, vadim.n.smelyanskiy@nasa.gov}
\begin{abstract}
The Sherrington-Kirkpatrick model with random $\pm1$ couplings is programmed on the D-Wave Two annealer featuring 509 qubits interacting on a Chimera-type graph.
The performance of the optimizer compares and correlates to simulated annealing. When considering the effect of the static noise, which degrades the performance of the annealer, one can estimate an improvement on the comparative scaling of the two methods in favor of the D-Wave machine. The optimal choice of parameters of the embedding on the Chimera graph is shown to be associated to the emergence of the spin-glass critical temperature of the embedded problem.
\end{abstract}
\maketitle

NP problems, such as classical paradigmatic computer science problems~\cite{lucas2014ising} as well as practical engineering problems~\cite{smelyanskiy2012near} can often be formulated efficiently as Quadratic Unconstrained Binary Optimizations (QUBO). These computational challenges can be seen as the task of finding the ground states of a disordered Ising spin glass often defined on a potentially highly-connected graph~\cite{fu1986application}.

One tantalizing approach to solve QUBOs in their Ising formulation is provided by programmable quantum annealing. While the founding principles of the technique have been investigated numerically~\cite{santoro2002theory}, analytically~\cite{morita2008mathematical} and experimentally~\cite{brooke1999quantum} in the last decade, the disordered, interacting, time-dependent and open nature of the many-body problem makes it very hard to draw universal conclusions about the power of the technique~\cite{zagoskin2014test}.

\begin{figure}
\includegraphics[width=\columnwidth]{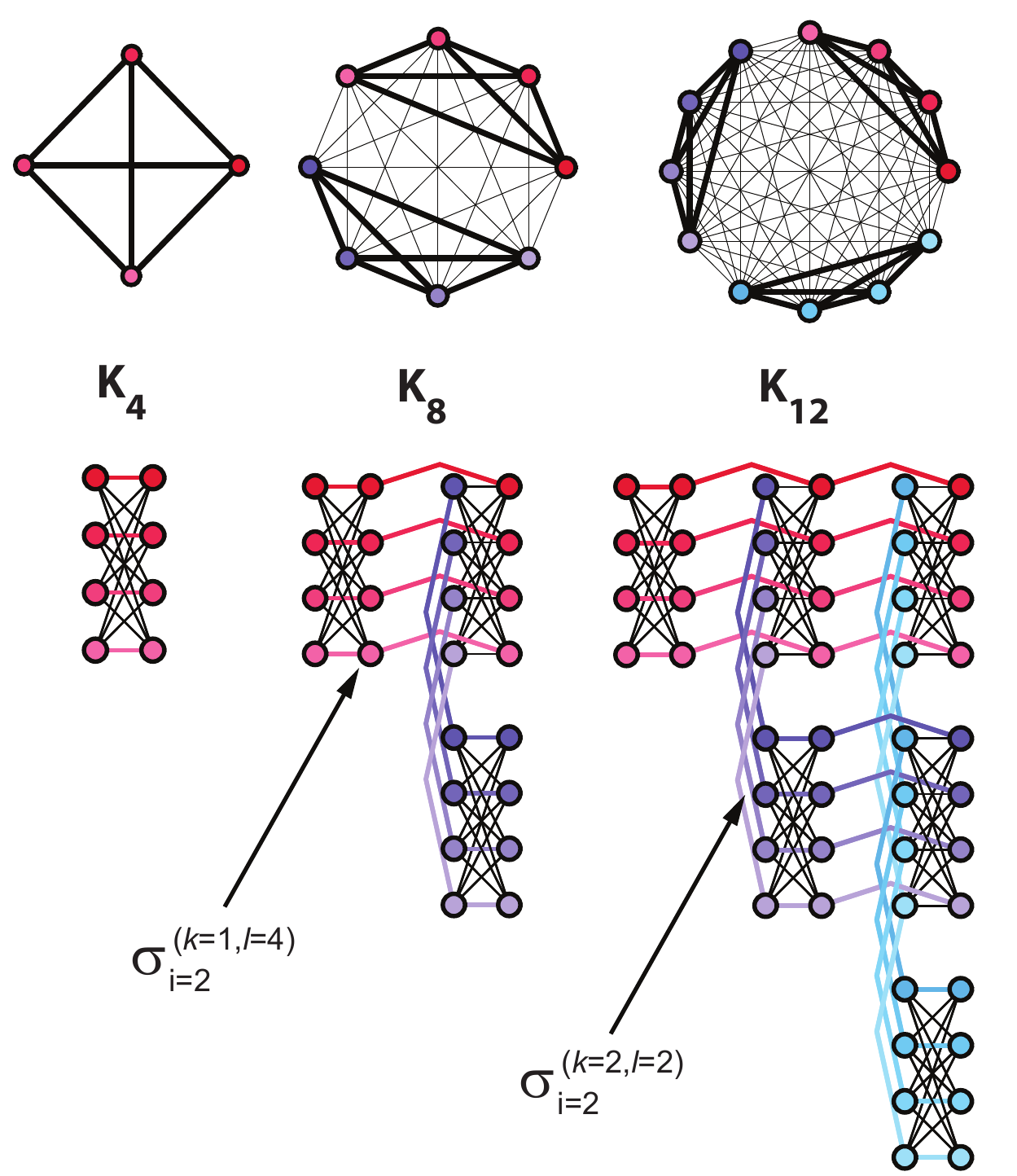}
\caption{(Color online) Illustration of the iterative embedding procedure of the SKM in the Chimera graph. Different colors represent the $N$ logical bits, which are arranged in $N/4$ groups of colors (reds, violets and cyans, indexed by $k$). The corresponding images of fully connected graphs on top show that logical bits in the same group of colors have two different ways to be connected by a physical coupling on the Chimera graph by having a thicker edge between them. The arrows indicate two qubits with their respective indices convention with reference to the labeling of Eq.~\ref{eq:embeddedH}, where $l$ indicates position within a given color group and $i$ is a running label of the position of the qubit in the chain starting from the uppermost (top-left) $i=1$ up to the last $i=N/4+1$.}
\label{fig:EmbeddingSK}
\end{figure}

One very recent development that boosted scientific activity in this field has been the commercialization of D-Wave Two optimizers, which implements the annealing approach by means of a solid state architecture consisting of hundreds of interlaced superconducting flux qubits~\cite{bunyk2014architectural}.
While the manufacturing methods and the computing technology are well documented, understanding the power of the machine is a formidable challenge for the aforementioned reasons, with the additional hindrance that the heavy integration of the circuitry entails the existence of static and dynamical sources of noise that are in part unknown.
For these reasons groups around the world have started to experimentally benchmark the machine~\cite{mcgeoch2013experimental,DashCPLEX,boixo2013quantum}, nurturing a lively discussion on whether the device is making functional use of quantum mechanics for computation~\cite{vinci2014distinguishing} and how to properly measure speedups between different computational/experimental algorithms~\cite{ronnow2014defining,smolin2013classical}.
On a more pragmatic level, the chip was also tested to evaluate its performance on toy-application problems in the fields of network diagnostics~\cite{APO2014QADMF}, artificial intelligence~\cite{planningquantum}, computational biology~\cite{perdomo2012finding}, mathematics~\cite{bian2012experimental} and machine learning~\cite{ogorman2014baysnet}. One typical occurrence in applied problems is when the QUBO to be solved is derived from a linear binary optimization problem with large number of constraints such as enforced equalities or inequalities between linear relations of variables. In this case the resulting penalty terms in the objective function form intersecting cliques whose minimization might be a hard computational problem for classical algorithms such as simulated annealing.

Motivated by the great value of quantifying the power of quantum optimization on valuable applications, in this work we report on the optimal programming guidelines and performance expectation of the D-Wave Two Vesuvius chip, applied to problems defined on fully-connected graphs with random couplings in the absence of longitudinal local fields. This Hamiltonian corresponds to the Sherrington-Kirkpatrick Model (SKM) with couplings randomized from a bimodal distribution of values $\pm 1$~\cite{PhysRevLett.35.1792}. The SKM is directly related to the graph partitioning problem~\cite{fu1986application} which is known to be NP-hard, and supports a spin-glass phase at finite temperature with transverse fields. For these reasons, it represents one of the most interesting benchmarks to evaluate the performance of the optimizer on structured problems.
Moreover, the encoding of the SKM on the D-Wave hardware has very interesting symmetry properties, allowing us to elegantly investigate general procedures common to all structured optimizations on annealers, such as the parameter setting of embedding and the error-correction, which in the general case require heuristic numerical pre-processing~\cite{APO2014DWtuning}.

The D-Wave Two "Vesuvius" chip hosted at NASA Ames Research Center features 509
working flux-qubits connected by 1455 tunable composite qubits acting as
Ising-interaction couplings~\cite{harris2010experimental}, arranged in a non-planar lattice known as a
"Chimera graph"~\cite{johnson2010scalable}.
In order to implement general Hamiltonians which are defined
on arbitrary graphs, it is customary to employ the graph-minor embedding~\cite{kaminsky2004scalable} technique.
This procedure consists of finding a set of connected subgraphs (logical bits or LB, corresponding to different colors in Fig.~\ref{fig:EmbeddingSK}) of the original graph such that
each LB can be associated to a node in the original graph. This association needs
to be such that for each two connected nodes there exists at least one edge between the qubits belonging to the associated LBs.
While the problem of finding an optimal (i.e. minimizing the number of required nodes) graph minor is itself NP-hard~\cite{eppstein2009finding} and is typically tackled with heuristic approaches~\cite{Cai-14}, for many graphs with a regular struture an efficient embedding can
be found systematically.

Fig.~\ref{fig:EmbeddingSK} shows an embedding of SKM in a triangular
portion~\cite{choi2011minor,klymko2013adiabatic} of the Vesuvius processor: each LB in the original problem of size N is
represented by (N/4)+1 qubits connected in a line. This means that this embedding procedure encompasses
an overhead of $N^2/4+N$ hardware qubits for encoding fully-connected graphs. Note that a quadratic
scaling of the embedding resources for SKM is expected for any hardware graph with fixed degree.
The embedding procedure is useful for the encoding of the problem Hamiltonian into the hardware
processor as long as the qubits in each LB are collapsed on the same z-value at the end of the annealing.  The basic idea is to ferromagnetically couple all qubits with a negative weight $J_{F}$ within a LB in such a way to energetically penalize discordant qubit states. The remaining couplings can be assigned to reflect the logical Hamiltonian of the problem to be solved.

With reference to the index convention illustrated in Fig.~\ref{fig:EmbeddingSK}, the actual Hamiltonian which is encoded in the annealing machine is then:

\begin{align}
  H_{\text{SKM}}=\left(A(t)\sum_{i}S_{i}^{X}+B(t)\sum_{ij}J_{ij}S^{Z}_i S^{Z}_j\right)\label{eq:SKH}\\
  \Longrightarrow A(t)\sum_{kli}\sigma^{x(kl)}_{i}+B(t)\sum_{kli}\bigg[-\sigma_i^{z(kl)}\sigma_{i+1}^{z(kl)}\label{eq:embeddedH}\\
  -\delta_{ki}\sum_{l^\prime<l}\frac{J_{(kl,kl^\prime)}}{J_{F}}\left(\sigma_k^{z(kl)}\sigma_{k+1}^{z(kl^\prime)}\right)\notag\\
  -\sum_{k^\prime = k+i}\sum_{l^\prime}\frac{J_{(kl,k^\prime l^\prime)}}{J_{F}}\left(\sigma_{k+1+i}^{z(kl)}\sigma_{k}^{z(k^\prime l^\prime)}\right)\bigg]\notag,
\end{align}

where $S^{*}_i$ and $\sigma^{*(kl)}_{i}$ are respectively the LB and the Pauli operators corresponding to the qubits along the $*$-direction. The logical SKM couplings $J_{ij} $ have been explicitly divided among the inter-cell couplings $J_{(kl,kl^\prime)}$ and the couplings between different groups of colors $J_{(kl,k^\prime l)}$ and the bounds on the summed variables are implied. Since the maximum allowed energy coupling in the D-Wave Hamiltonian is 1, increasing $J_{F}$ is equivalent to rescaling the logical couplings. A(t) and B(t) are the time-dependent coefficients that define the annealing schedule performed by the machine~\cite{johnson2011quantum}. It is immediately apparent that from the dynamical perspective that the optimal prescription on the value of $J_{F}$ might be tricky to evaluate despite the fact that it is always possible to set its magnitude to be sufficiently high to make sure that the target ground state still lies at the bottom of the embedded classical spectrum~\cite{choi2008minor}.

\begin{figure}
\includegraphics[width=\columnwidth]{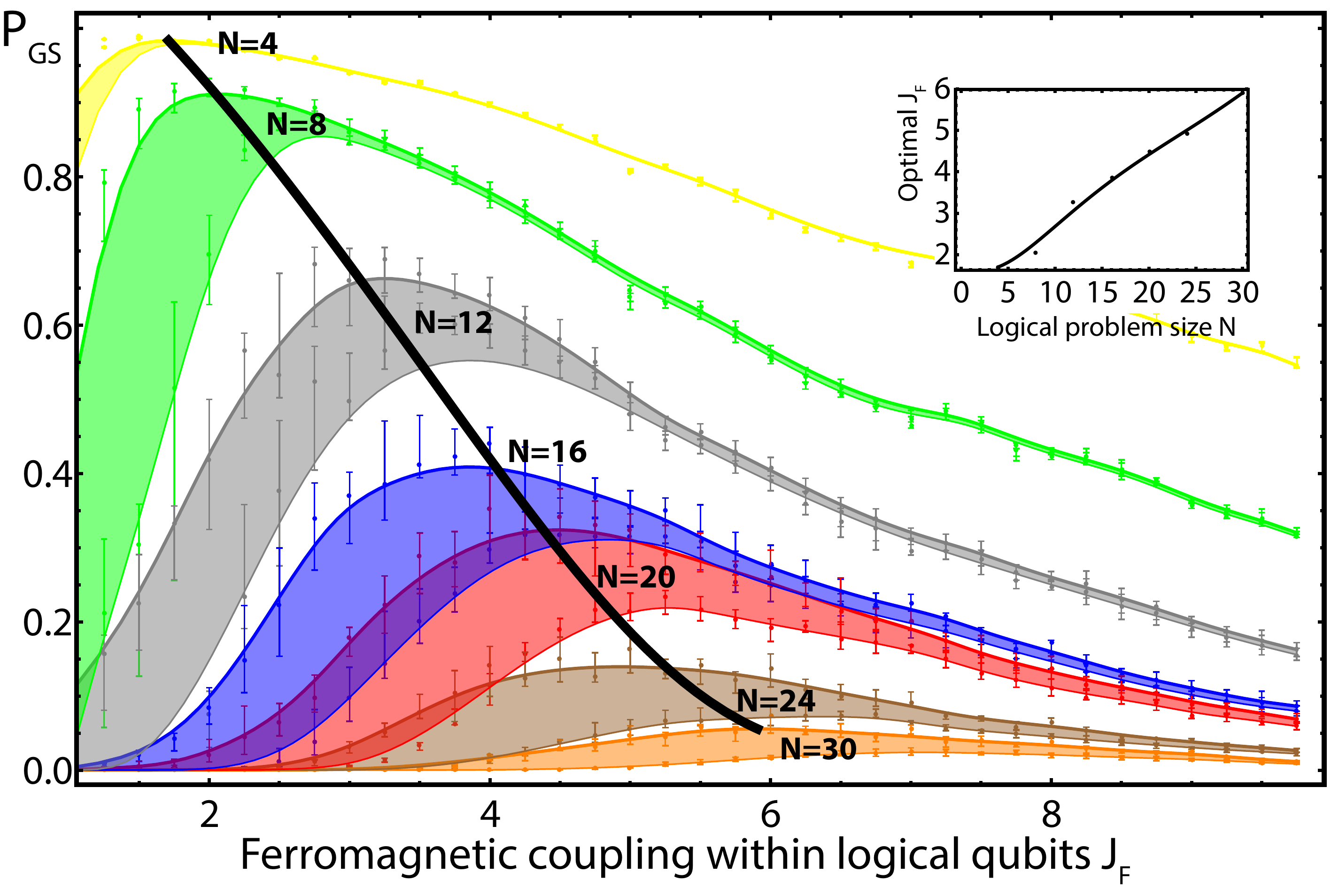}
\caption{(Color online) Bezier fit of median probabilities of finding the ground state $P_{\text{GS}}$ of the encoded SKM (independently checked with exact enumeration code) after 400000 runs on the D-Wave Two machine for every problem instance. The average is taken over 80 instances per size, and runs are performed using 10 random gauges~\cite{APO2014DWtuning}. The black line and the inset indicates the optimal $J_{F}$ for a given size, which increases with $N$ as a power law close to $\sqrt{N}$. Errorbars are obtained through resampling.}
\label{fig:JFERROplotOPT}
\end{figure}

Fig.~\ref{fig:JFERROplotOPT} shows the median probability $P_{GS}$ for the analog optimizer (run at fixed annealing time $\tau$=20 $\mu s$) to reach the ground state. For a given problem size N it depends significantly on $J_{F}$ and goes to zero for large and small value of $J_{F}$. For $J_{F}\simeq 1$, the ferromagnetic couplings are not energetically stronger than the logical couplings and we expect that the problem is not well encoded. Indeed many chains representing LBs are found in excited states (i.e. having 1 or more kinks) as illustrated by the colored bands of the plot which displays the improvement on $P_{GS}$ obtained by a post-processing procedure which tries to recover logical states from broken chains by doing majority voting (similarly to error-correction/repetition codes~\cite{error-corrected-lidar,young2013adiabatic}\footnote{If we randomize the logical state of the LB in case of presence of kinks, instead of using the majority voting rule, we still recover a comparable amount of solutions. Nevertheless the majority voting rule is more effective, suggesting that the defects which are created at the boundaries of the chain of qubits don't propagate far away before being frozen (towards the end of the annealing)}). Conversely, for sufficiently large $J_{F}$, defects in the LBs are suppressed, but the overall annealing success probability decreases after an optimal $J_{F}$. The appearance of this maximum can be connected to the expectation that the annealing dynamics is more efficient when the ferromagnetic LBs become correlated at the same time as the described SKM enters the spin-glass phase. This is because once the chains feel the ferromagnetic fixed point (for a transverse field of $A(t)\simeq B(t)$) their dynamics slows down and might reasonably impede the development of correlations between the logical states, while in the paramagnetic state they are more easily subject to the formation of kinks. This argument is also supported by the scaling of the optimal coupling, which can be fit as a power law with an exponent close to $1/2$, which is consistent with the critical transverse field of the embedded SKM, which goes proportionally to $B(t)\sqrt{N}/J_{F}$ ~\cite{suzuki2013quantum} (connected to Fig.~\ref{fig:spinglassT} described later on). Comparisons with embedding and runs on embedded 2D-lattices also support the above theory (as detailed in the Supplemental Material (SM)).


\begin{figure}
\includegraphics[width=\columnwidth]{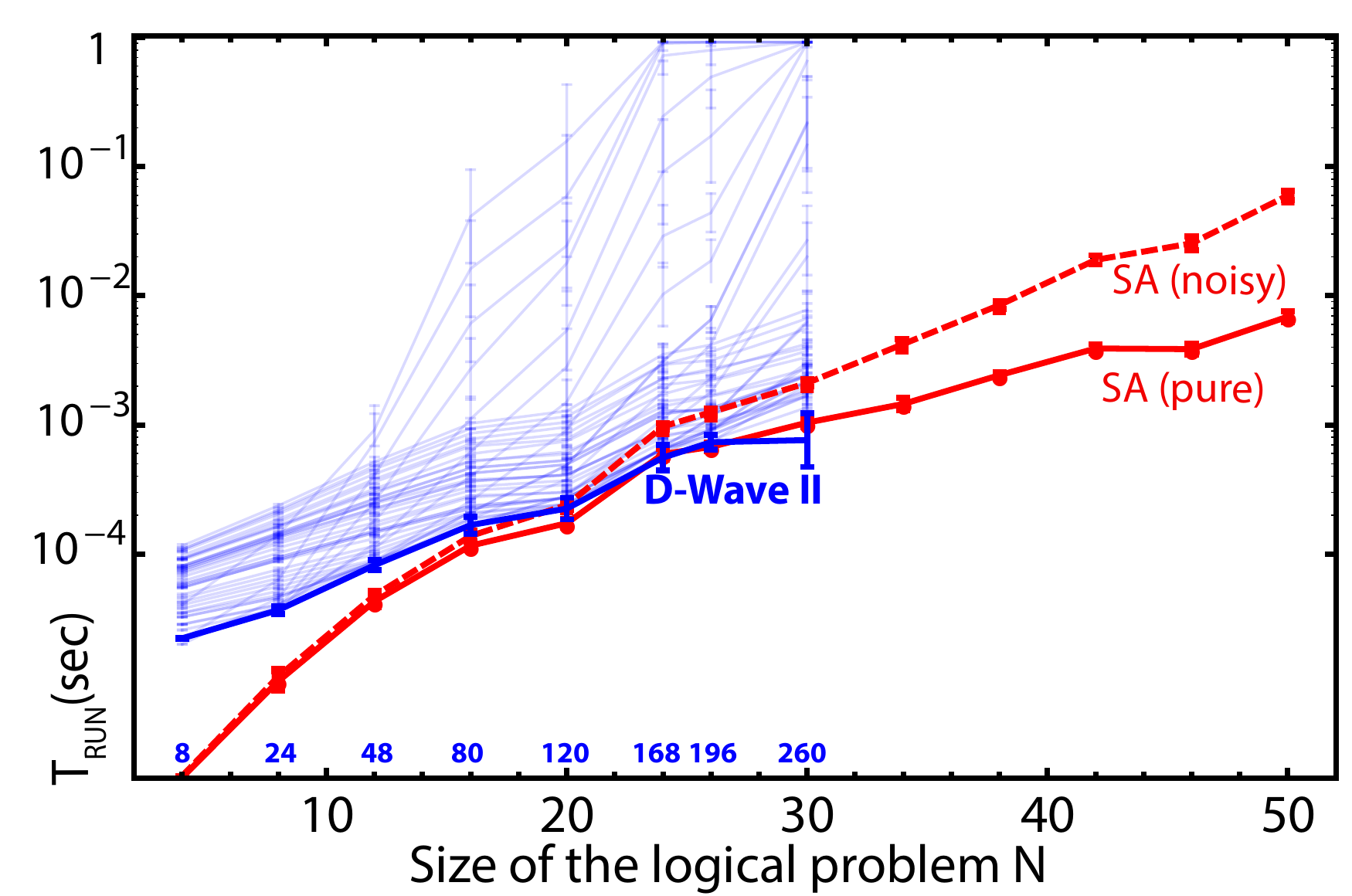}
\caption{(Color online) Blue curves: Each curve is the expected D-Wave median runtime for achieving 99\% probability of finding the ground state ($P_{\text{GS}}$), computed as $T_{\text{RUN}}=\tau R$, with $R$ being the expected number of repeated annealing cycles $R=\log(0.01)/\log(1-P_{\text{GS}})$, at fixed $J_{F}$. The result shown includes the error-correction procedure, whose processing time is not considered. The best possible result irrespective of $J_{F}$ for each problem size is highlighted in the thicker blue line. Red curves: simulated annealing results for the bare logical problem (thick) and with the introduction with the respective noise model of the D-Wave machine described below.}
\label{fig:JFERROplot}
\end{figure}

Figure \ref{fig:JFERROplot} shows the median expected runtime (in seconds) $T_{\text{RUN}}$ for the annealing device to find the ground-state with 99\% probability, for different $J_{F}$ and the experimentally shortest possible $\tau=20 \mu s$.  The thicker blue line shows the scaling of complexity with the system size assuming the optimization on $J_{F}$ with a precision of $\Delta J=0.25$. This exponential scaling and the absolute runtime seems very similar (and correlates well~[SM]) to the performance of simulated annealing (SA) on the same $logical$ instance set extended up to $N=50$ whose runtime is optimized over $\tau$ measured on Intel Xeon E5-2680v2 processors. However, it is well known that the Hamiltonian parameters programmed on the analog optimizer are subject to low-frequency noise that can be modeled as static gaussian disorder realization for each instance~\cite{trevor}. One could argue that this noise (whose presence is not fundamental but rather an engineering issue) introduces an artificial handicap in the evaluation of the performance of the D-Wave machine, as the programmed problem might significantly differ from the target objective function to be minimized. We introduced the noise effect in the logical instance runs with SA in order to compare the scalings with on fair grounds.  While on the scale of the maximum physical energy programmed in the problem Hamiltonian (i.e. 3.2 GhZ) this model of noise has a negligible effect~\cite{trevor}\footnote{For this reason we run SA on logical problems rather than embedded problems.}, the rescaling of the absolute energy of the logical parameters due to the introduction of $J_{F}$ proportionally amplifies the relevance of the unwanted disorders. The considered noise model spoils the $J_{(kl,k^\prime l^\prime)}$ couplings of Eq.~\ref{eq:embeddedH} and introduces artificial longitudinal local fields \footnote{There is also a random artificial second nearest neighbor coupling due to qubit imperfections, but this contribution goes to zero in the classical limit.}.
More specifically, as the logical couplings $J_{ij}$ are chosen to be $\pm1$, this implies that the problem Hamiltonian to be compared with D-Wave II runs at fixed $J_{F}$ must be spoiled as follows:
\begin{align}
H_{\text{dev}}=H_{\text{SKM}}+\left[\sum_{ij}\xi_{J}^{ij}
S_{i}S_{j}+\sum_{i}\xi_{h}^{i}S_{i}\right],\label{eq:noisy}
\end{align}
where $\xi_{J}^{ij}$ and $\xi_{h}^{i}$ are disorder realizations with gaussian distribution around zero of respective standard deviations $\sigma^\xi_J$=0.035 and $\sigma^\xi_h$=0.05~\cite{trevor}.  Results are averaged over 1000 realizations for every instance and new optimal speeds have been computed for the final scaling~[SM]. What is observed is that starting from N=12 the noise significantly affects the probability for the spoiled system to find the ground state of the ideal Hamiltonian. As detailed in the SM, for every fixed level of noise $\propto J_{F}$ there is indeed a problem size above which the noise tends to shift the ground state of the noisy Hamiltonian outside the manifold of the degenerate ground states of the ideal Hamiltonian, independently from the algorithm used to compute the ground state. We note that this effect is likely to be dominant over the slow-down of the LB dynamics conjectured to be responsible of the sharp decrease in performance of the device for large $J_{F}$ observed in Fig.\ref{fig:JFERROplotOPT}, and more analysis is needed to establish if this is the case.

While unsurprisingly the current limitations on the number of qubits do not allow us to draw final conclusions on whether the machine has a sound speedup with respect to classical digital methods, the scaling results are encouraging. While it is now established that speedup might emerge artificially due to suboptimal annealing speed~\cite{ronnow2014defining} ($\tau=20\mu s$ would supposedly become optimal only for larger N) as well as due to correlation between different subsequent runs~\cite{boixo2014evidence}, we have shown evidence that this is likely to be masked by the detrimental effect of the noise (which is expected to be significantly reduced in future generations of the device~\cite{trevor}). Most importantly, our work elucidates how evaluating the comparative performance of analog optimization with respect to algorithmic methods on necessarily embedded problems is more delicate than it is on natively structured problems. This is largely because the correct representation of the target problem requires an optimal tuning of the analog optimizer, which is dependent on the hardware architecture and the programmability precision. The statistical reasonings behind benchmarks~\cite{ronnow2014defining} and complexity estimation~\cite{glassychimera} on natively structured problems performed by previous works need to be extended considering that in embedded problems the number of LBs does not reflect the number of qubits for the comparison of required resources~\footnote{As an example, we note that due to the unparallelizable nature of SA simulations on SKM, limited quantum speedup as definite in~\cite{ronnow2014defining} should take into account that also classical computational resource availability scales with N, differently from benchmarks in finite-connectivity lattices.}.
Moreover, the fact that the logical Hamiltonian is emergent from a corse-graining of the hardware Hamiltonian, which has ferromagnetic correlations due to embedding, carries with it potentially profound consequences regarding the expected complexity of the annealing procedure on the logical problem. In the SKM this also means that the shape and the location of the critical region associated with the spin-glass phase is dependent on the internal representation parameters such as embedding topology and optimal $J_{F}$.

In order to gain insights on these issues, in Fig.~\ref{fig:spinglassT} we examined, by means of SA simulations, the emergence of the spin-glass phase of the embedded SKM model (Eq.~\ref{eq:embeddedH}), i.e. the appearance of a pseudo-critical (normalized) spin-glass temperature $T_{\text{SG}}$ as a function of $\alpha_F=J_{F}/\sqrt{N}$. Our findings are compatible with a scaling exponent $\kappa\simeq3$ (as expected from the exact SKM) for the universal spin-configuration overlap Binder ratio behavior $g\simeq G[N^{1/\kappa}\alpha_{F} (T-T_{\text{SG}})]$~\cite{glassychimera}[SM],  and with an increase of the pseudo-critical temperature with $\alpha_F$ towards the theoretical value of the unembedded model, which in the thermodynamic limit is $T_{\text{SG}}=1/\alpha_F$ (but for finite $N$ the Binder curves intersect at a smaller value~\cite{bhatt1985search}).  This means that embedded problems can belong to a different universality class than random chimera problems, answering a question at the center of the current discussion in the quantum annealing community~\cite{glassychimera}.

\begin{figure}
\includegraphics[width=\columnwidth]{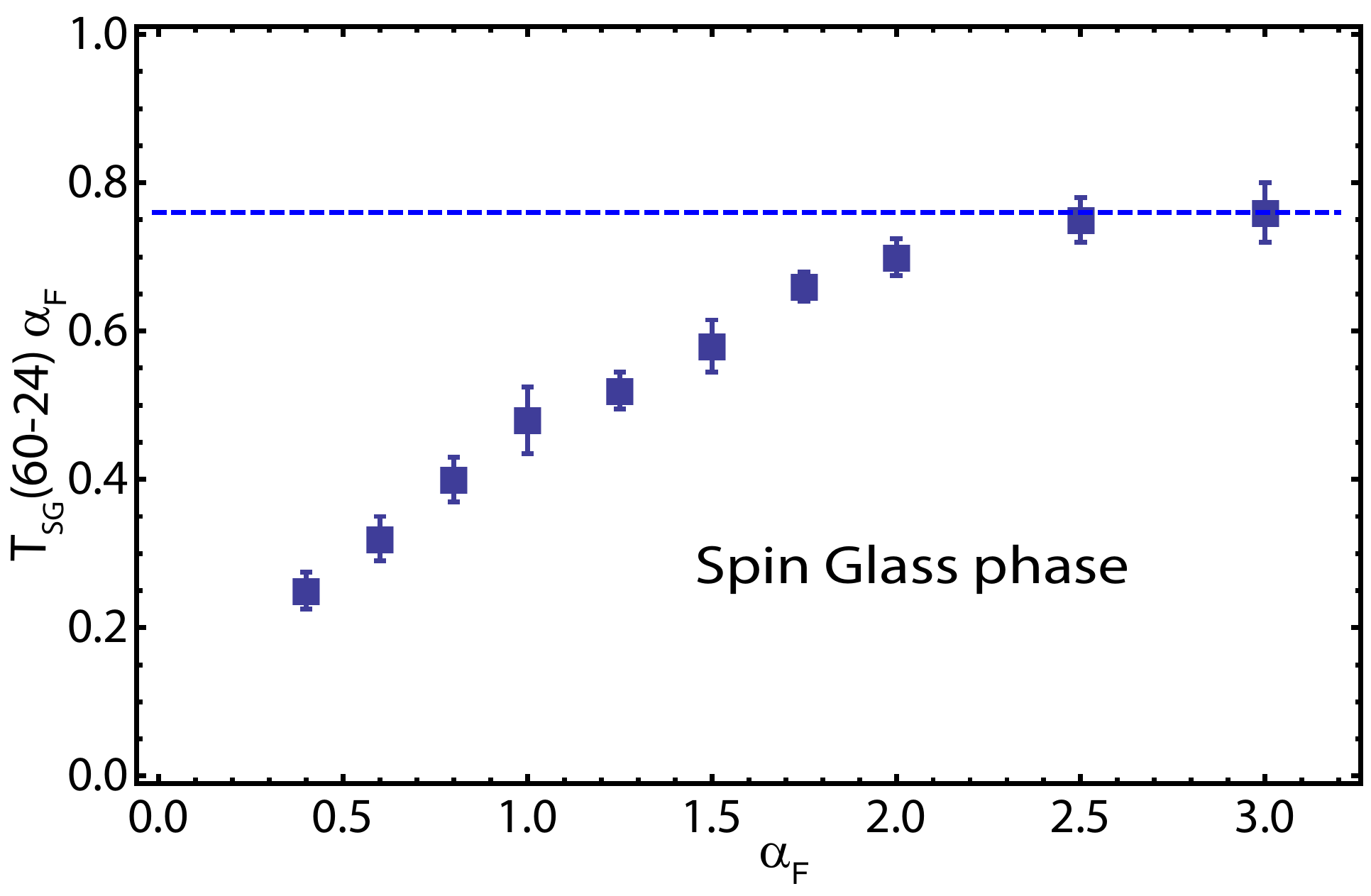}
\caption{(Color online) Binder ratio intersections for embedded SKM in Chimera. The approximation of the critical temperature has been computed as the intersection of the $N=60$ and $N=24$ curves $g(T)=\frac{1}{2}\left[3-\langle q(T)^{4}\rangle/\langle q(T)^2\rangle^2\right]$ of spin configuration overlaps $q=\frac{1}{N}\sum_{i}S^A_{i}S^B_{i}$ between two replicated runs $A$ and $B$ at various temperatures $T$ for each $\alpha_F$ in the figure. Error bars correspond to error-propagation over the intersecting region, and the dashed black line indicates the intersection relative to the logical problem ($T_{SG}(60-24)$ saturates in the limit $\alpha_F \rightarrow \infty$ to a value of $(0.76\pm0.08)/\alpha_F$, which is smaller than the known exact value of 1/$\alpha_F$. This discrepancy is due to finite size effect for the ensemble sizes studied in this work. See SM for more details.}
\label{fig:spinglassT}
\end{figure}

These results support the intuition that the ferromagnetic couplings need to increase as $\sqrt{N}$ (up to logarithmic corrections) in order to properly represent the SKM for large sizes. Interestingly, the experimentally optimal $\alpha_{F}$ (see Fig.~\ref{fig:JFERROplot}) in our runs is close to 1.0, meaning that the machine is better off optimizing the spectrum of an embedded representation of the SKM model whose critical temperature is appreciable. Moreover, the critical temperature explored by the D-Wave machine at optimal parameter setting is larger than the experimental one, which might have profound consequences on the asymptotic computational complexity of quantum annealing on the embedded SKM.

\begin{acknowledgments}
We acknowledge useful discussions with Dr. Trevor Lanting, Dr. Alejandro Perdomo-Ortiz, Dr. Eleanor G. Rieffel, Prof. M. Troyer and Prof. H. Katzgraber. S. Mandr\`a was supported by NASA (Sponsor Award Number: NNX14AF62G). \end{acknowledgments}

\bibliographystyle{unsrt}

\pagebreak
\begin{widetext}
\begin{center}
\textbf{\large Supplemental Material: \\Quantum optimization of fully connected spin glasses}
\end{center}
\end{widetext}
\subsection{\normalsize{Optimal parameter for the classical simulated annealing}}

In order to compare the performances of the D-Wave II device with respect to
other classical methods, we studied the probability of success of simulated annealing (SA)
heuristics \cite{kirkpatrick1983optimization} on the same
instances that we run on the annealing machine. Since classical algorithms are not limited by hardware connectivity,
we performed the classical simulations using
the ``logical'' Hamiltonians, namely the original problem Hamiltonian without any
further embedding. The number of logical spins for the Sherrington-Kirkpatrick model (SKM)
considered varies from $N = 4$ to $N = 60$, with $N = 30$ the upper limit
of the maximum number of spins that can be actually embedded on a 512-spins Chimera
Hamiltonian without breaking symmetries.

As ``fair'' quantity to compare the performances of both the quantum (D-Wave II) and
classical (SA) devices, we used the ``expected'' runtime defined as
\begin{equation}\label{eq:expected_run_time}
	T_{\exp}(m) = m\, \tau(m) \frac{\log(1-s)}{\log(1-p(m))},
\end{equation}
with $m$ the number of sweeps and $p(m)$ the success probability after $m$ sweeps, while
$s$ is the probability of success that one wants to achieve (from here $s = 0.99$)
~\cite{ronnow2014defining}. Here $\tau(m)$ is the cpu time it takes for a classical
core to perform a single sweep.
The ``optimal'' computational time $T_{\text{opt}}$ will be defined
as
\begin{equation}\label{eq:eff_runtime}
	T_{\text{opt}} = \min_m T_{\exp}(m),
\end{equation}
where $m^* = \text{argmin}_m T_{\exp}(m)$ is the optimal number of sweeps, or ``speed''.

The SA heuristic depends on two parameters: The initial temperature $T_0$ and the total
number of sweeps $m$. At the beginning, the configuration is initialized to
a high temperature configuration. Starting from an initial temperature $T_0$
we slowly cool down the system until $T = 0$ is reached. The cooling down
is performed using a linear schedule in $m$ sweeps. In the case of SKM,
the initial temperature is chosen equal to the critical spin glass temperature
of the model, aka $T_0 = \sqrt{N}$ \cite{thouless1977solution}.
For any $T_0$, $m$ and instance, we repeated SA schedules 1000 times.

\begin{figure}[t!]
	\centering
	\includegraphics[width=0.50\textwidth]{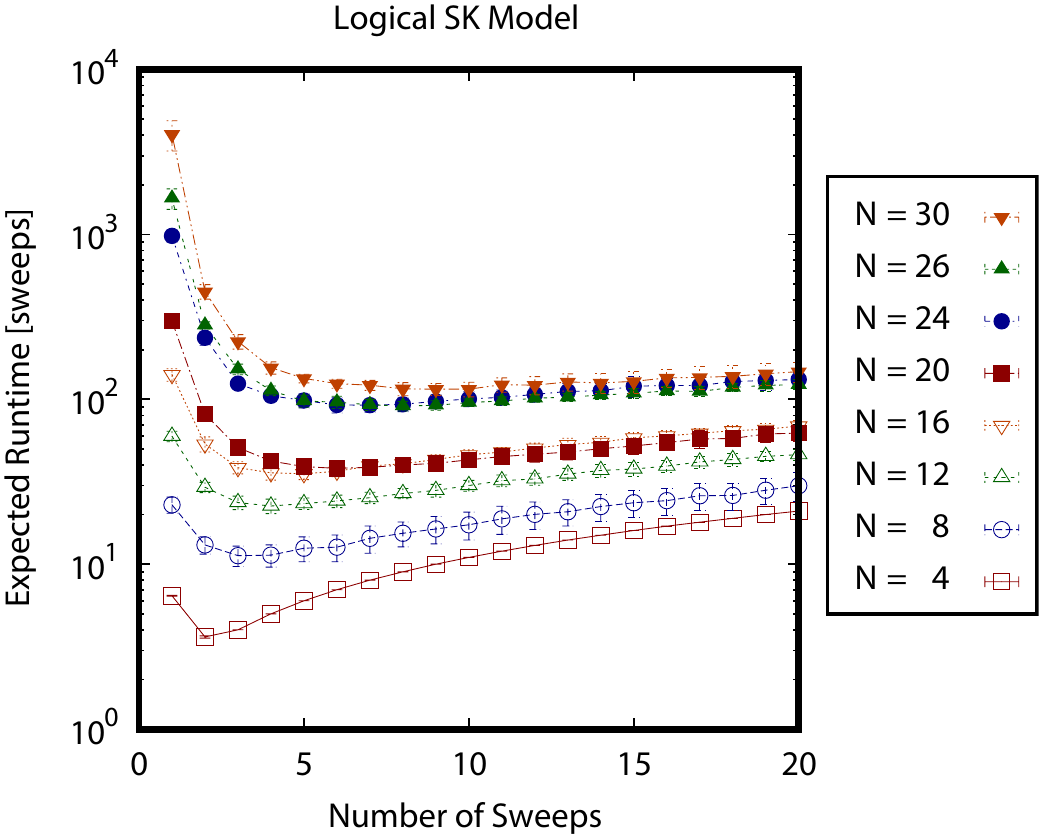}
	\caption{\label{fig:SK_exp_runtime}
		\textbf{Expected computational time $T_{\exp}$ as in Eq.~(\ref{eq:expected_run_time}),
		by varying the number
		of sweeps for the logical SK model, at different number of logical spins $N$.}
		The optimal number of sweeps (speed) is defined as the number of sweeps which minimizes
		$T_{\exp}$.}
\end{figure}

Figure (\ref{fig:SK_exp_runtime}) shows the median expected runtime by varying the number of
sweeps for SKM.
The optimal computational time $T_{\text{opt}}$ can be easily
identified as the minimum of those curves.


\subsection{\normalsize{Role of the noise for the classical simulated annealing}}

The DWave II runs are plagued by uncontrollable noise originated by the non ideality of manufactured qubits, the low-frequency fluctuations, the unwanted flux offsets, the low-frequency noise, on-chip crosstalk as well as by errors associated to the many digital-to-analog converters.
\begin{figure*}[t]
	\centering
	\includegraphics[width=0.48\textwidth]{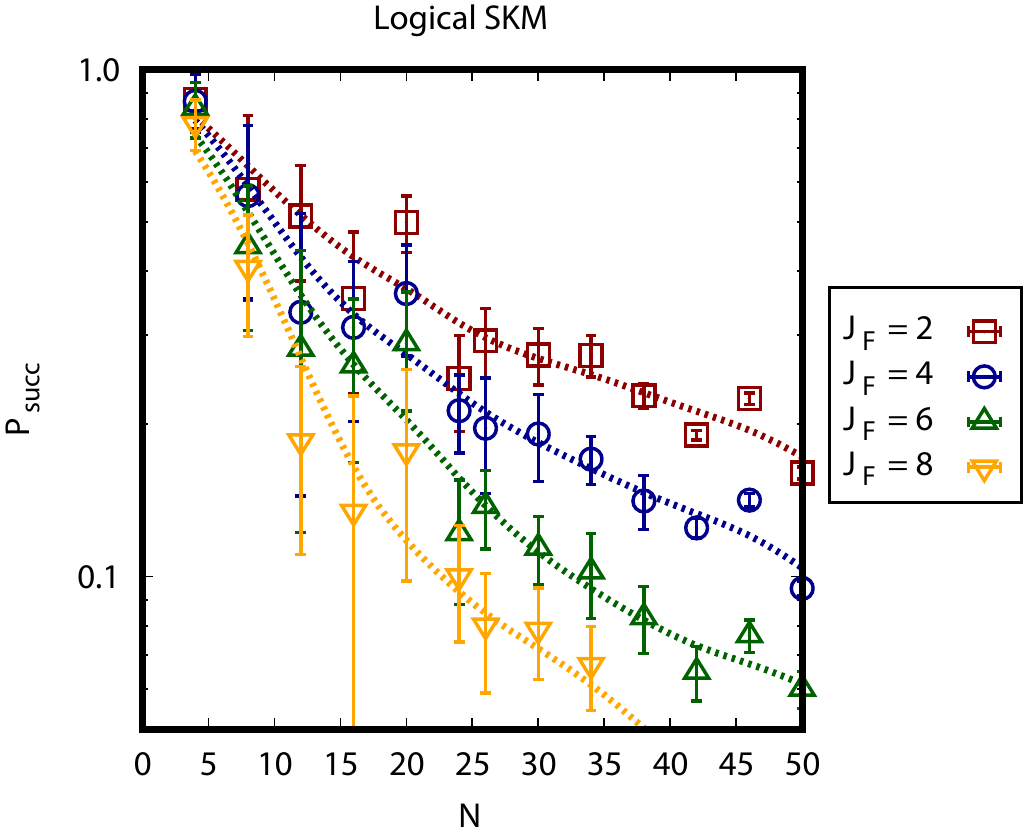}
	\hspace{0.1cm}
	\includegraphics[width=0.48\textwidth]{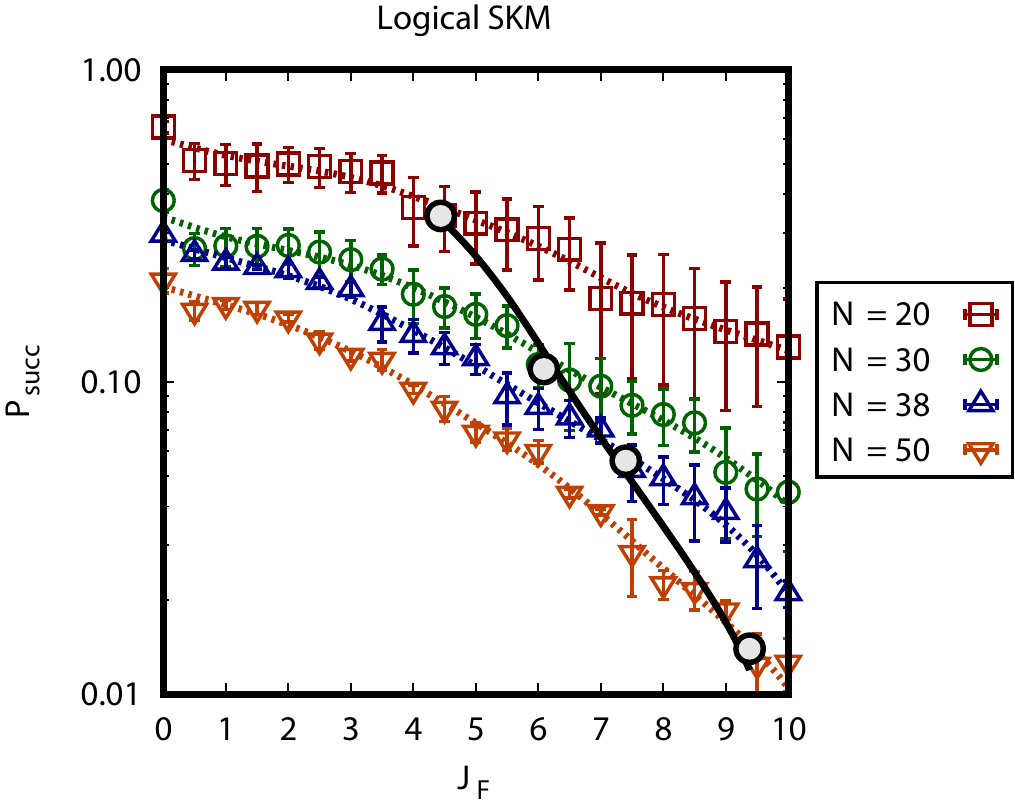}
	\caption{\label{fig:SK_noise}
		\textbf{Noise quickly degrades the performance of the classical SA.}
			Figures show respectively (in log scale) the probability of success by varying the number of logical spins
			$N$ (Left), the probability of success by varying $J_F$ (Right).
			Data are optimized for any $N$ and $J_F$, averaging over 200 disorder realization.
			Black line indicates the optimal $J_F$ for the D-Wave II device.
		}
\end{figure*}
\begin{figure}[!t]
	\centering
	\includegraphics[width=0.4\textwidth]{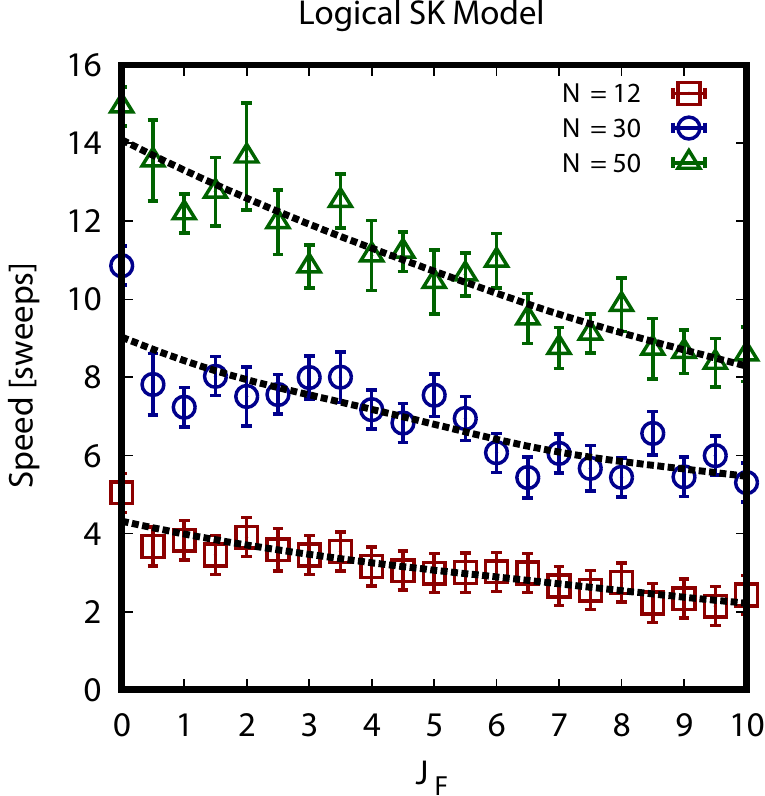}\\
	\vspace{0.2cm}
	\includegraphics[width=0.4\textwidth]{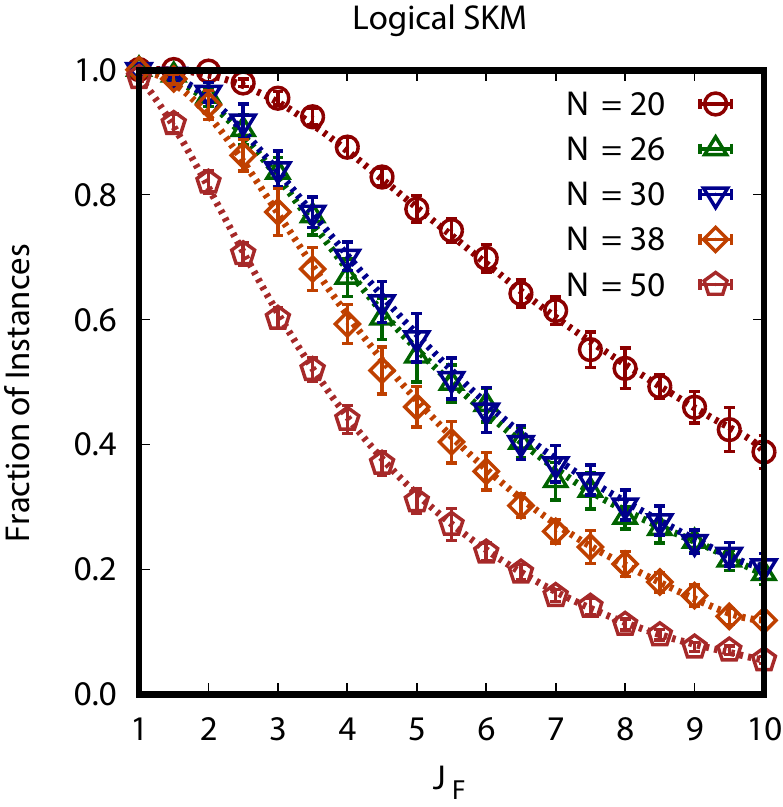}

	\caption{\label{fig:optM_SK_noise}
		\textbf{Optimal number of sweeps (speed) decreases by increasing \boldmath{$J_F$}}.
			Top Panel: optimal number of sweeps (speed) by varying $J_F$, at different
			number of logical spins $N$.			
			Bottom panel: the fraction of spoiled Hamiltonians
            (200 different realizations of disorders) for which the ground state corresponds to
			the ground state of the unspoiled logical Hamiltonian.	These curves are independent on the annealing speed as long as
            the schedule reaches the ground state with any finite probability.	}
\end{figure}
In this paper, we used the simplest model where noise on couplings
and local field are uncorrelated and Gaussian distributed (see
Eq.~(3) of the main text), with
respectively $\sigma^\xi_J = 0.035$ and $\sigma^\xi_h = 0.050$, from D-Wave measurements~\cite{trevor}, expressed in units of the maximum energy scale $J_{max}\simeq3.2 GHz$.
This noise is expected to be present even if couplings and external fields are set to zero.
To understand the effects of the noise on classical annealers, we
created a spoiled Hamiltonian by adding uncorrelated noise to all
couplings and to all local fields of random instances of the SKM Hamiltonian, Eq.~(1)
of the main text.

Since the optimal $J_F$ for the D-Wave device scales with the number of
logical spins, as reported in Fig.~(2) of the main text,
larger instances will be noisier than smaller instances.
Therefore, to correctly reproduce the effects of the noise
on the classical SA, in the logical spoiled SK Hamiltonian the noise
is chosen to be proportional to $J_F$, i.e.
\begin{subequations}\begin{align}
	\sigma^\xi_J &= 0.035\,|J_F|\\
	\sigma^\xi_h &= 0.050\,|J_F|
\end{align}\end{subequations}

The effects of the noise on the performance of SA
are shown in Fig.~(\ref{fig:SK_noise}), where
data has been obtained by using the optimal number of sweeps,
considering the effect of the noise,
for any fixed number of logical spins $N$ and $J_F$.
%
Black line
on the right panel indicates the optimal $J_F$ of the D-Wave II device (see Fig.~(2) of the main text).
As one can see, the performances of the classical SA quickly drop by increasing $J_F$.
Observe that the probability of success of the spoiled SKM for $N = 30$ is almost half
than the probability of success of the unspoiled SKM.

Speeds for the spoiled Hamiltonian are displayed in the top panel of Fig.~(\ref{fig:optM_SK_noise}):
Interestingly, the speed decreases by increasing the noise. This scaling is in accordance
with the fact that it is unlikely that the spoiled and the unspoiled Hamiltonian share
the same ground state for large noise, as depicted in the bottom panel of Fig.~(\ref{fig:optM_SK_noise}),
so that larger annealing times actually reduce the performance of SA in the presence of noise.

\subsection{\normalsize{Correlation plots of the effective computational times}}

\begin{figure}[t]
	\centering
	\includegraphics[width=0.4\textwidth]{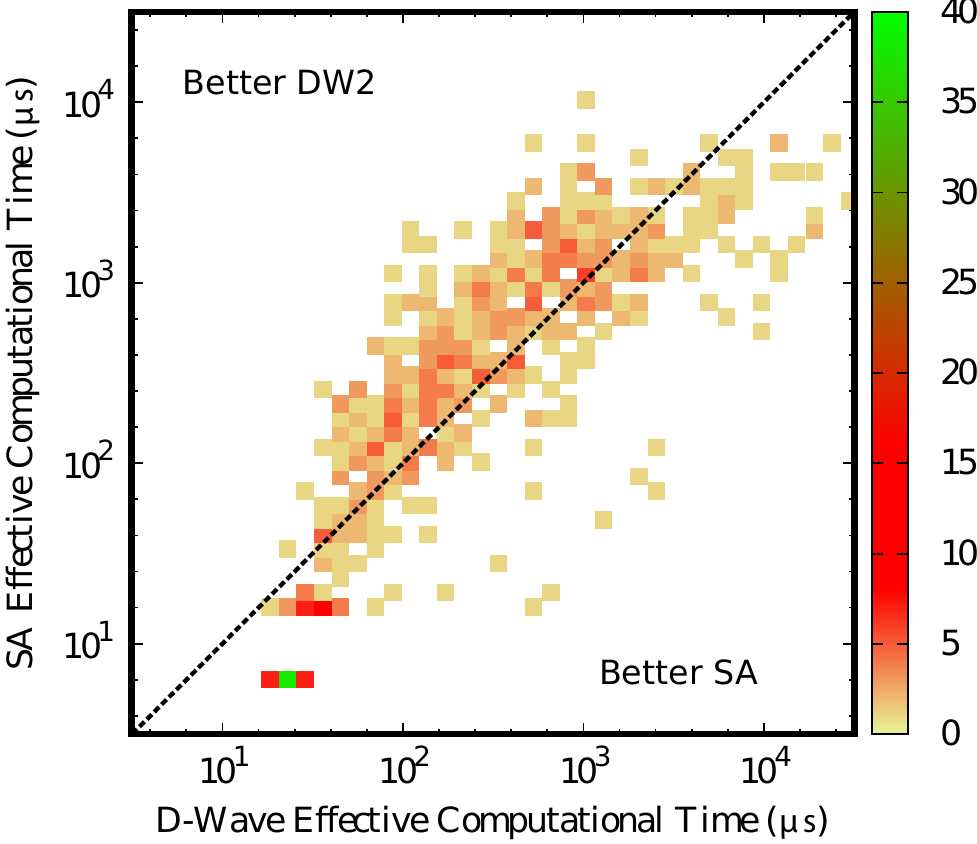}

	\vspace{0.2cm}
	\includegraphics[width=0.4\textwidth]{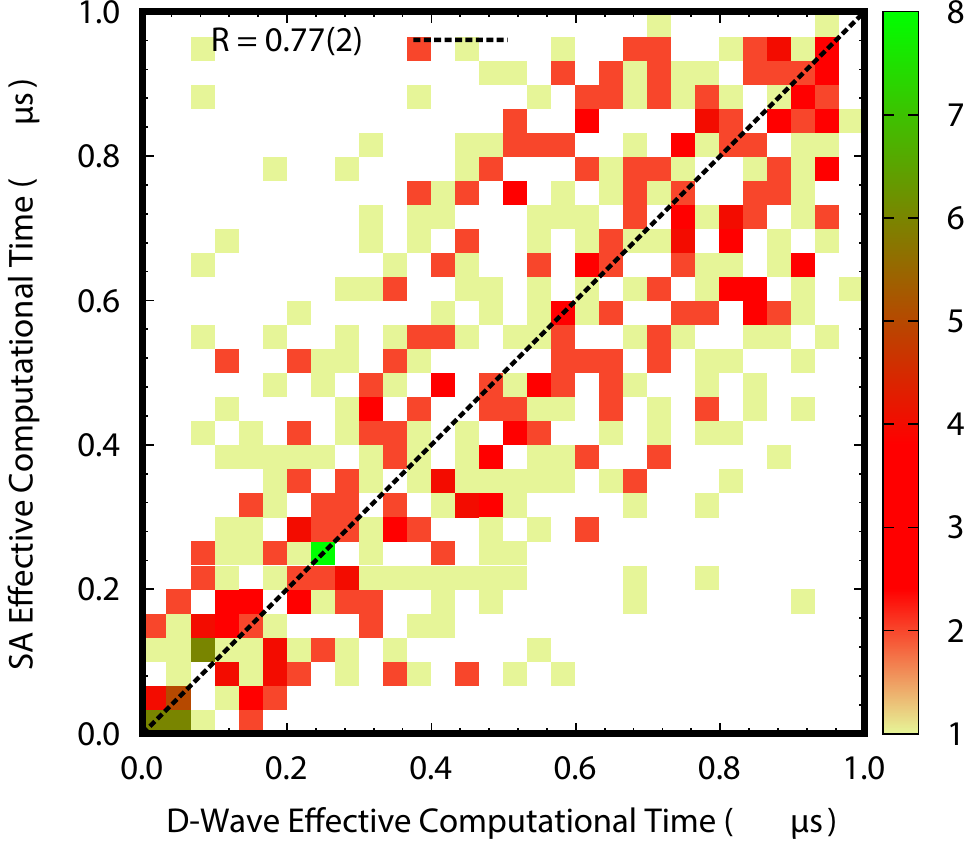}
	\caption{\label{fig:corr_SA_DW2}
		\textbf{The effective computational times of both the D-Wave II (DW2)
			device and the classical
			simulated annealing (SA) are comparable}. Correlation plots of the effective
			computational time between the D-Wave II device and the classical simulated annealing
			either using a direct comparison (Left) or the copula (Right).
			For the analysis, all the instances regardless the number of logical spins
			$N$ are used. Colors indicate the number of instances.			
			Although the D-Wave
			II device has a slightly better performance than the classical simulated annealing,
			the correlation coefficient $R$ is rather high.
		}
\end{figure}

In this section we compare the median optimal computational time $T_{\text{opt}}$ in SA simulations, as defined in Eq.~(\ref{eq:eff_runtime}), with the median optimal runtime of the D-Wave II (DW2) device considering optimization over $J_{F}$ and fixed annealing time of $\tau=20\mu s$, on the noise-free problem. To compare DW2 and SA, we use either a direct comparison of $T_{\text{opt}}$ instance by instance
or the ``copula'' of $T_{\text{opt}}$, namely the correlation of the ordered rankings~\cite{ronnow2014defining} of the value of $T_{\text{opt}}$,
which are respectively the left panel and the right panel of Fig.~(\ref{fig:corr_SA_DW2}).
The linear coefficient $R$ is computed as:
\begin{equation}
	R = \frac{\text{Cov}(n,n^\prime)}{\sqrt{\text{Var}(n)\text{Var}(n^\prime)}},
\end{equation}
with $n$ and $n^\prime$ the rank positions of the $T_{\text{opt}}$ for the two different
devices. The D-Wave
II device has a slightly better performance than the classical simulated annealing
for instances which requires larger annealing time, even if
the correlation coefficient $R$ is rather high ($R \approx 0.77$).
%

\subsection{\normalsize{Calculation of the critical temperature}}

\begin{figure*}
	\centering
	\includegraphics[width=0.32\textwidth]{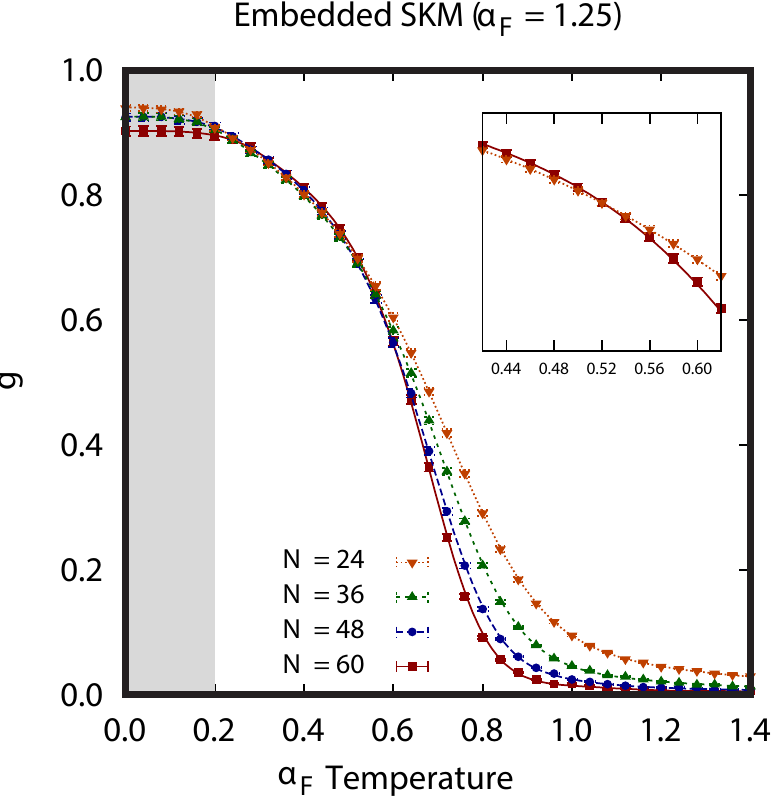}
	\hspace{0.1cm}
	\includegraphics[width=0.32\textwidth]{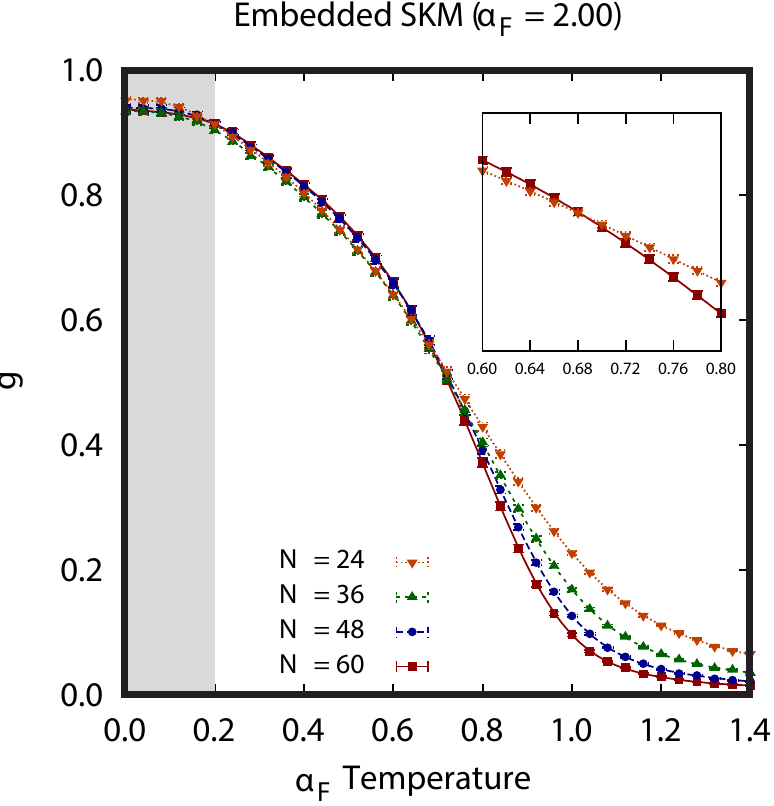}
	\hspace{0.1cm}
	\includegraphics[width=0.32\textwidth]{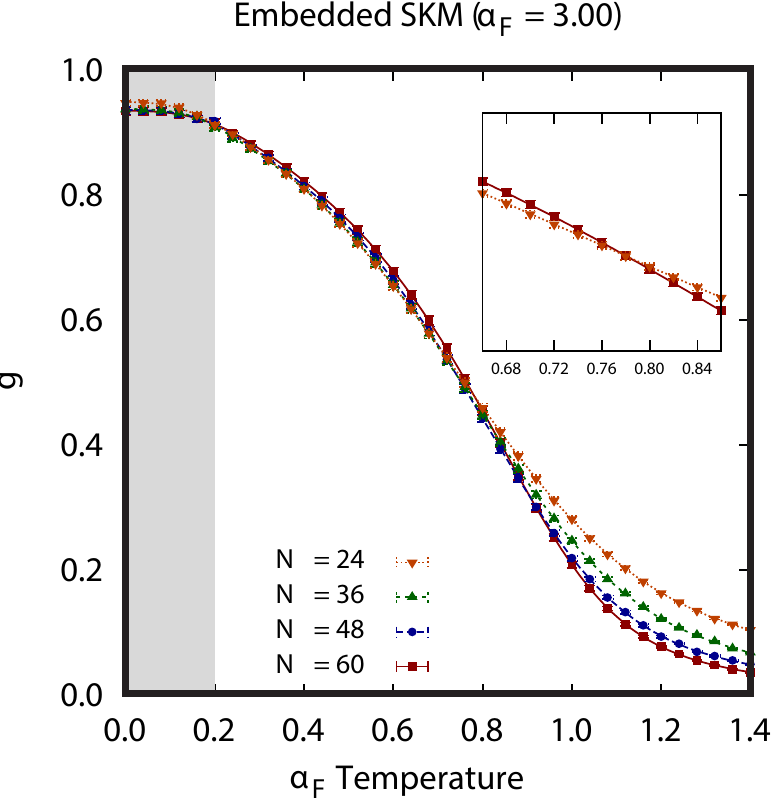}
	\caption{\label{fig:binder}
			\textbf{At fixed \boldmath{$\alpha_F$}, Binder ratio
			curves $g$ have an intersection, which indicates that the system
			undergoes to a spin glass phase transition at low, but non zero, temperature.}
			Figures show different Binder ratios $g$ by varying the number of SK logical spins
			$N$ embedded into
			the Chimera graph for $\alpha_{F} = 1.25$ (Left), $\alpha_{F} = 2$
			(Center) and
			$\alpha_{F} = 3$ (Right). Shaded area indicates the region where
			the fully equilibration is not reached. As well as the logical SKM which has
			a critical spin glass temperature $T_c = 1$, the embedded SK Hamiltonian
			undergoes to a spin glass phase transition at low temperature,
			as indicated by the intersection of the Binder ratios $g$.
			}
\end{figure*}
\begin{figure*}
	\centering
	\includegraphics[width=0.32\textwidth]{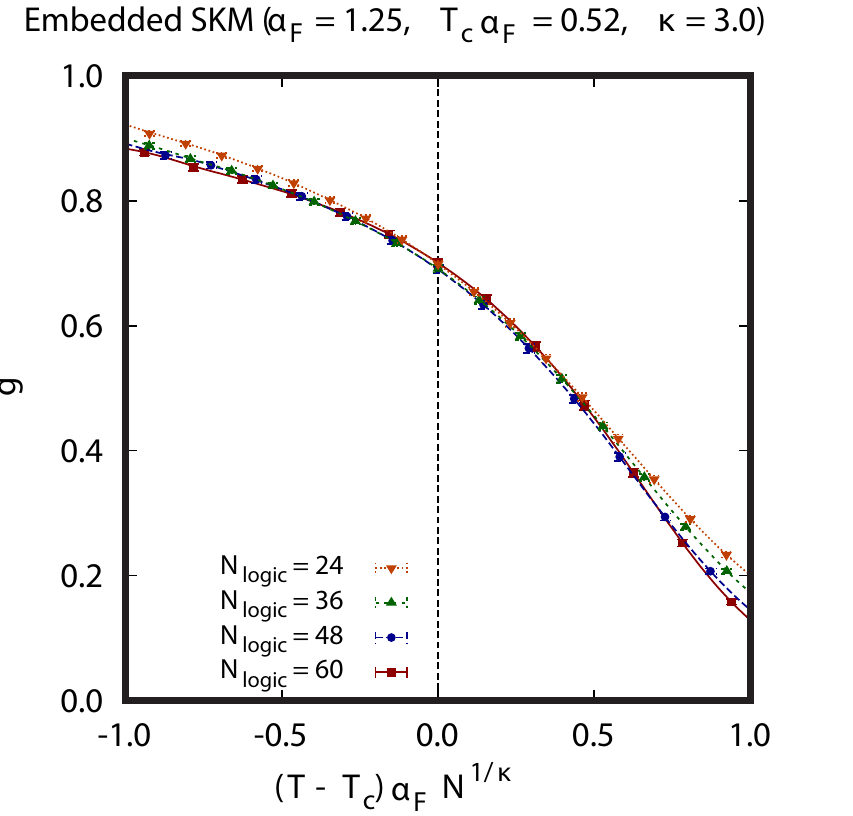}
	\hspace{0.1cm}
	\includegraphics[width=0.32\textwidth]{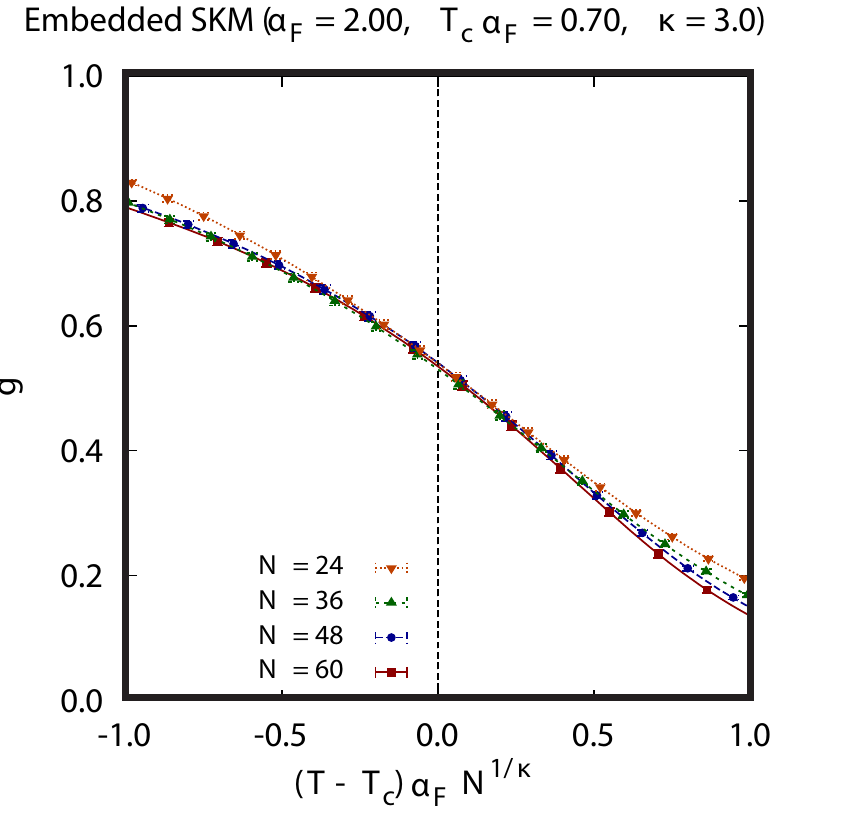}
	\hspace{0.1cm}
	\includegraphics[width=0.32\textwidth]{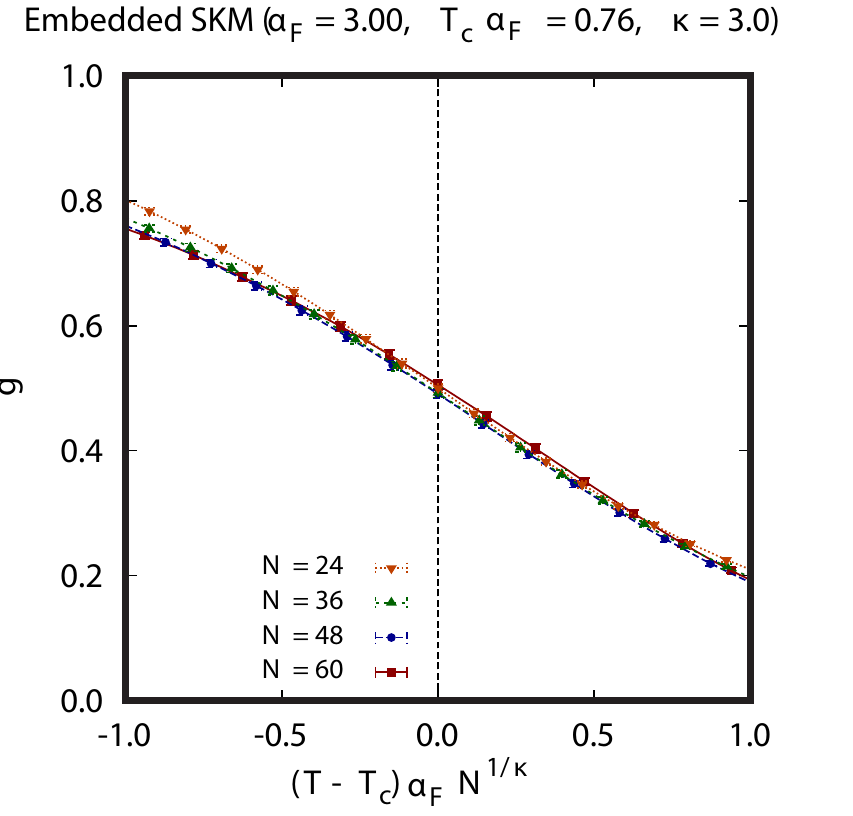}
	\caption{\label{fig:binder_scaling}
			\textbf{The critical exponent \boldmath{$\kappa$} for the embedded SK
				Hamiltonian is compatible with $\kappa = 3$, the same as the logical SKM}.
				Finite size scaling of the Binder ratio $g$ by varying the number
				of logical spins $N$.
				The different curves of $g$ scales very well around $T = T_c$ when
				$\kappa = 3$ is chosen.
			}
\end{figure*}
Spatially random systems can undergo a spin-glass phase transition where local metastable states
dominate
the thermodynamics of the system. The classical ($B(t)=0$) SKM was one
of the first model for which the existence of the spin glass phase transition has been proven
\cite{thouless1977solution}
to occur at
$T_c = |J|\sqrt{N}$, where $|J|^2$ is the variance of the random couplings.
As exploited in Ref.~\cite{edwards1975theory,sherrington1975solvable,fischer1975static}
the spin glass phase transition in a system with N spins can be detected using the
spin configuration overlap defined as
\begin{equation}
	q = \frac{1}{N}\sum_i \sigma_i^{(1)}\sigma_i^{(2)},
\end{equation}
where $\sigma^{(1)}$ and $\sigma^{(2)}$ represent two independent replicas with
the same disorder. In the case of the embedded SKM, the replicas refer to the actual physical spin configurations of the embedded SKM Hamiltonian, and the number of spins is accordingly chosen to be the total number of physical spins used. In the high temperature limit, where
the system is in its paramagnetic phase, the overlap distribution $P(q)$
follows a Gaussian distribution of width $\sqrt{N}$, where $N$ is
the number of spins. On the contrary, in the spin glass phase,
$P(q)$ converges to a distribution with a non trivial support \cite{parisi1987spin}.

In order to detect a spin glass phase transition for the embedded SKM Hamiltonian
(Eq.~(2) of the main paper) we numerically studied the Binder
ratio of the spin configuration overlap
\cite{binder1981critical,bhatt1985search,katzgraber2014glassy}
\begin{equation}
	g(T) = \frac{1}{2}
	\left[
		3 - \frac{\left\langle q^4\right\rangle}{\left\langle q^2\right\rangle^2}
	\right],
\end{equation}
where $\left\langle\cdot\right\rangle$ denote both the statistical
mechanics average at fixed temperature $T$ and the disorder average $J$.
The Binder ratio $g$ is a dimensionless parameter defined so that $g\to0$ in the
paramagnetic phase and $0\leq g \leq 1$ in the spin glass phase,
observing a scaling $g \sim G \left[N^{1/\kappa}\left(T - T_c\right)\frac{J_F}{\sqrt{N}}\right]$,
 in terms of $T_c$ (the critical temperature for which the system undergoes to
a spin glass phase transition) and $\kappa$ (the critical exponent corresponding to
2-$\alpha$, standard notation~\cite{dutta2011transverse}, by means of Josephson's identity of hyperscaling).

In Fig.~(\ref{fig:binder}), we show the Binder ratio $g$ computed for the
embedded SKM Hamiltonian in Eq.~(2) of the main text, by varying number
of logical spins $N$ for the embedded SK Hamiltonian at fixed
$J_F = \alpha_F \sqrt{N}$, where $\alpha_F$ is chosen respectively $\alpha_F = 1.25$
(left panel), $\alpha_F = 2.0$ (middle panel) and $\alpha_F = 3.0$ (right panel).
As one can see, for sufficiently large $\alpha_F$ the curves of $g$ show an intersection
in $T = T_c$, which indicates the presence of a spin glass phase transition.
Figure~(4) of the main text shows $T_c$ computed as the intersection of the Binder
ratios $g$ for $N = 60$ and $N = 24$ logical spins. Since the full
equilibration is obtained only for $T > 0.2$ (see next Section for more details),
only points for reliable intersections are shown. However, we cannot
exclude \emph{a priori} the existence of intersections of the Binder ratio
for small $\alpha_F$ at very small temperature.
Figure~(\ref{fig:binder_scaling}) shows the same data presented in Fig.~(\ref{fig:binder})
but properly rescaled, in order to better appreciate the critical temperature
$T_c$ and the critical exponent $\kappa$.

\begin{figure*}
	\centering
	\includegraphics[width=0.5\textwidth]{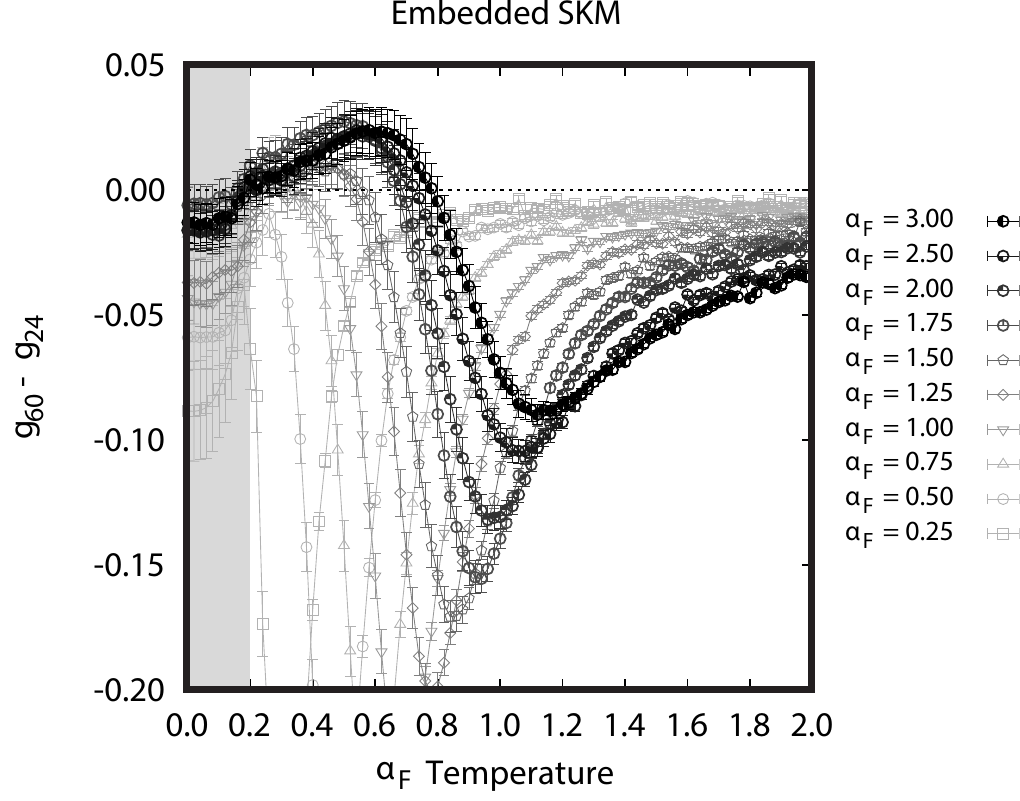}
	\hspace{0.1cm}
	\includegraphics[width=0.36\textwidth]{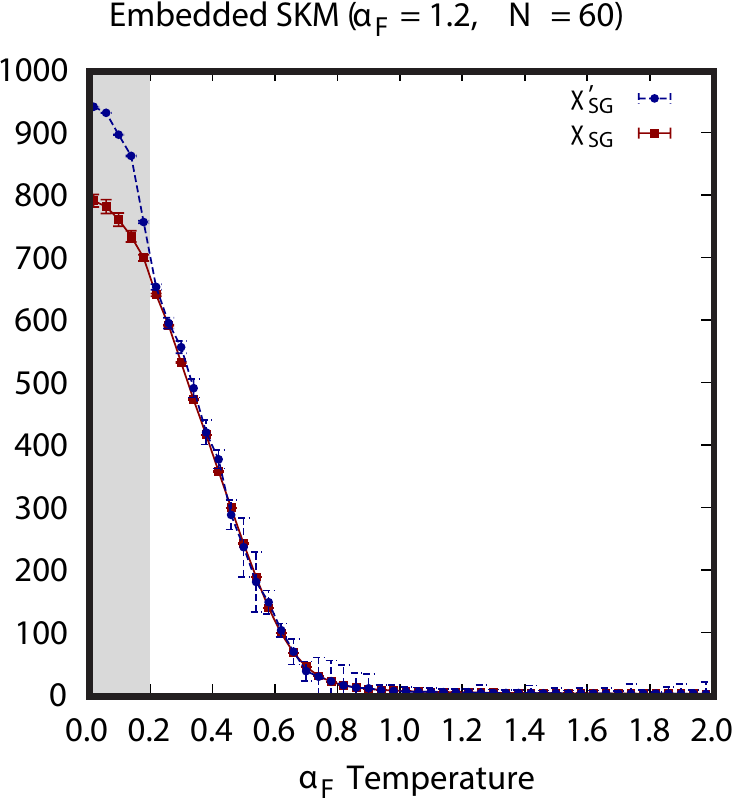}
	\caption{\label{fig:diff_binder}
			\textbf{Binder ratio curves \boldmath{$g$} show an intersection
				for sufficiently large $\alpha_F$, which indicates the presence of a
				spin glass phase transition.} Left: Difference of the Binder ratios $g$ computed
				for $60-24$ logical spins, by
				varying the temperature and $\alpha_{F}$.
				For sufficiently large $\alpha_{F}$, the Binder ratios show
				an intersection in a specific critical temperature $T_c$,
				where a spin glass phase occurs.
				Shaded area indicates where the fully equilibration is not reached.
                Right:
			Spin glass susceptibility computed using both
			the lower bound estimation $\chi_{\text{SG}}$
			(red circle line) and the upper bound estimation $\chi^\prime_{\text{SG}}$
			(blue circle line) of the true spin glass susceptibility.
			The shaded area indicates where $\chi_{\text{SG}}$ and $\chi^\prime_{\text{SG}}$
			diverge. The curves
			merge only for $T > 0.2$, where a full equilibration is guaranteed.
			}
\end{figure*}

Noteworthy, the curves of the Binder ratio $g$ scale well around $T_c$
when $\kappa = 3$ is chosen as critical exponent, which is the same
critical exponent $\kappa$ of the logical SK model
\cite{parisi1993several,aspelmeier2008finite,billoire2011finite}.
Finally, Fig.~(\ref{fig:diff_binder}) shows the difference of the Binder ratios
using $N = 60,\,24$ (left panel) and $N = 60,\,36$ (right panel).
The intersection with the zero axis are used to compute $T_c$ as a function of
$\alpha_F$, as show in Fig.~(4) of the main text.

\subsection{\normalsize{Equilibration of the embedded SK Hamiltonian}}

As described in the main text, the embedding of the SKM creates
long ferromagnetic chains of ``physical'' spins, which correspond to a single ``logical''
spin (LB). When the intra-chain ferromagnetic coupling $J_F$ becomes comparatively very large, the whole
chain behaves like a true logical spin and then, we expect for the embedded SKM
to show the same thermodynamics properties of the logical SKM.
\begin{figure*}
	\centering
	\includegraphics[width=0.425\textwidth]{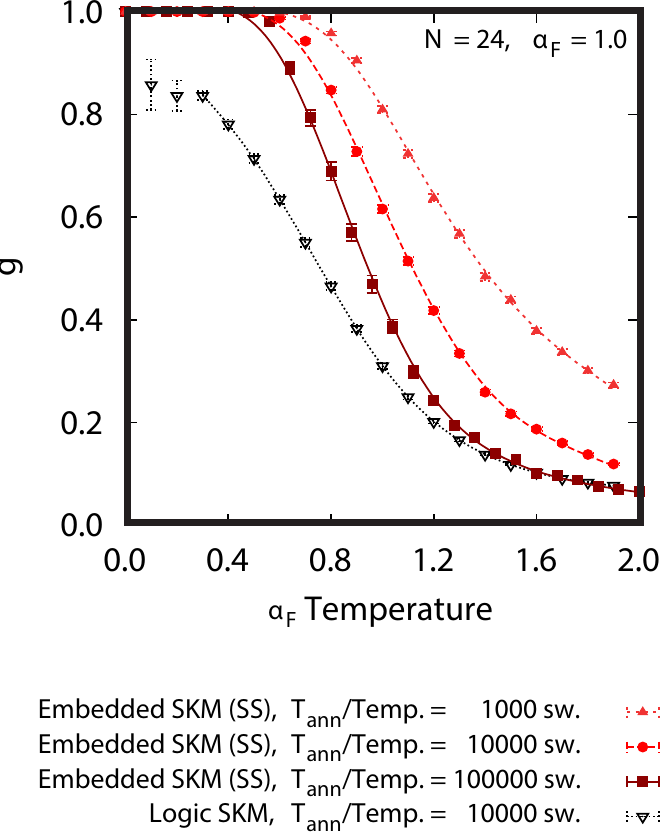}
	\hspace{0.1cm}
	\includegraphics[width=0.45\textwidth]{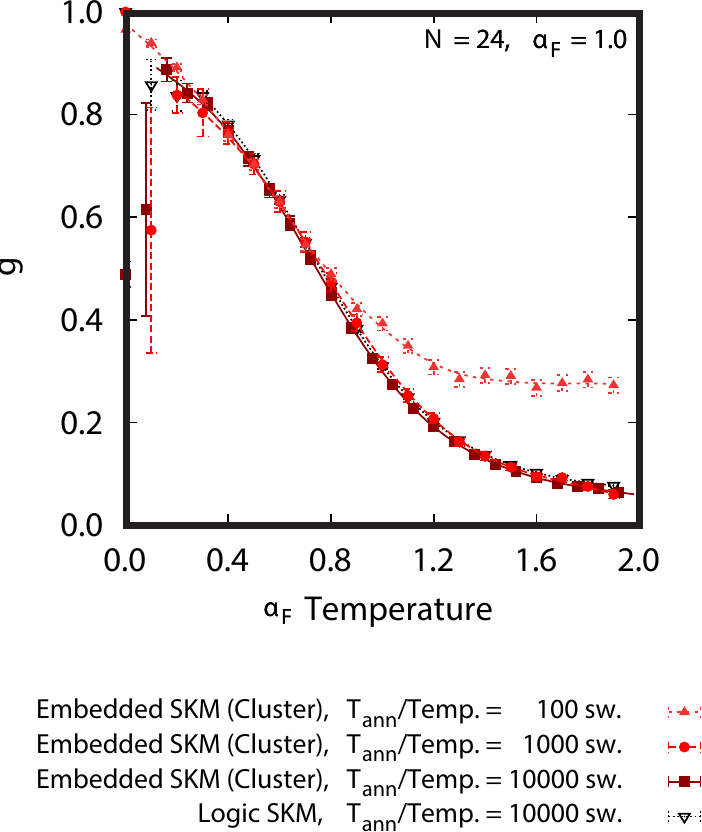}
	\caption{\label{fig:equilib_SK}			
			\textbf{Single spin update (SS) requires a longer equilibration time with respect
			to our modified cluster (Cluster) update.} Figures show the Binder ratio $g$
			by varying the temperature for both the single spin update and our
			modified cluster update for the embedded SK Hamiltonian
			using $\alpha_F = 1.0$.
			}
\end{figure*}
Unfortunately, due to the presence of these long ferromagnetic chains,
equilibration of the embedded SK Hamiltonian happens to be extremely long if the
a standard single spin flip Metropolis-Harris is used to perform SA simulations.
Indeed, for low temperature and $J_{F} \ll |J_\text{ij}|$, where $J_{ij}$ are
the couplings or the original logical SK model, spins belonging to the same chain prefer
to stay aligned. Since the probability to create a defect in a polarized chain
is proportional to $\exp(-2\beta |J_{F}|)$, large part of the equilibration time
is spent to try to flip a whole chain. Left panel of Fig.~(\ref{fig:equilib_SK}) shows
the Binder ratio $g$ for the embedded SK Hamiltonian (solid symbols)
computed by using a Metropolis-Harris single spin-flip update, by varying
the number of equilibration sweeps $\mathcal{T}_{\text{eq}}$ for each temperature.
Curves are compared with the Binder ratio $g$ for the logical SKM.
As one can see, the correct equilibration is obtained only for large temperature,
above the spin glass critical temperature $T_c = 1$.

In order to reach a faster equilibration for the embedded SKM,
we propose a variant of the Wolff cluster method
\cite{wolff1989collective} (a generalization
of the original Swendsen-Wang cluster method \cite{wang1990cluster}),
which takes into account the existence of the logical superstructures.
In the Wolff cluster method, a cluster is created at any time step
by using the following rules:

\begin{enumerate}
	\item Chose a random spin $i$ which represents the ``center'' of the cluster $C$.
	\item Create clusters: neighbors of the center are included as members of the cluster with a probability:
			$$
				p(\sigma_i,\,\sigma_j) =
					1 - \exp(-\beta|J_{ij}|+\beta J_{ij}\sigma_i\sigma_j).
			$$
			New neighbors are considered if that particular pair were not considered before.
	\item The creation process continues until no new neighbors are added.
	\item Flip the whole cluster.
\end{enumerate}
Given a cluster $C$ created by using the above rules,
the transition probability from a given configuration of the system $\sigma$ to a configuration
where the cluster $C$ is flipped can be written as \cite{wang1990cluster}
\begin{align}
	W(\sigma\to\sigma^\prime,\,C) &= w_\text{bulk}(\sigma,\,C)
		\prod_{\left\langle i,\,j\right\rangle\in\partial C}\left[1-p(\sigma_i,\,\sigma_j)\right]\nonumber\\
		&= w_\text{bulk}(\sigma,\,C)\,
		e^{-\beta\sum_{\left\langle i,\,j\right\rangle\in\partial C}(|J_{ij}|-J_{ij}\sigma_i\sigma_j)}.
\end{align}
In the above equation, $\partial C$ indicates the ``border'' of the cluster $C$, namely those spins
inside the cluster $C$ that share a coupling with spins outside $C$, and
$w_\text{bulk}(\sigma,\,C)$ is the probability
to create the ``bulk'' of the cluster $C$ without its border. Since $w_\text{bulk}(\sigma,\,C) = \sigma(\sigma^\prime,\,C)$,
because all the spins in the bulk of the cluster are flipped at the same time,
the detailed balance
$W(\sigma\to\sigma^\prime,\,C)\,p(\sigma) = W(\sigma^\prime\to\sigma,\,C)\,p(\sigma^\prime)$,
with $p(\sigma) = e^{-\beta H(\sigma)}/\mathcal{Z}$, is trivially satisfied.

As described in \cite{wang1990cluster,wolff1989collective},
the Wolff cluster method works well in the presence of
many domain-walls: in this case, flipping clusters reduces the equilibration time in
simulated annealing simulations by quickly removing borders between two neighbors clusters
with opposite sign. Unfortunately, the Wolff cluster method perform poorly
for fully connected spin-glass model like the SK model, since the Wolff procedure
typically creates clusters which contain almost all the spins. Therefore, the
Wolff procedure cannot be used ``as it'' on the embedding SK Hamiltonian since
we expect the same thermodynamics properties as the logical SKM for large $J_F$.

\begin{figure}[t]
\centering
\includegraphics[width=0.45\textwidth]{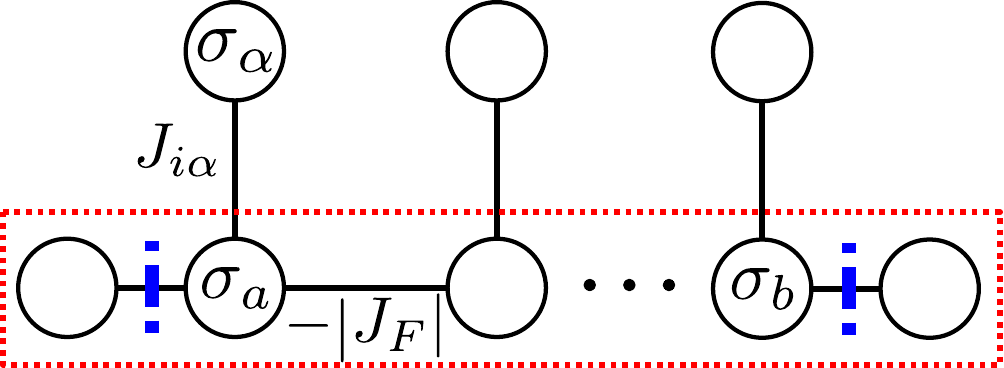}
	\caption{\label{fig:simple_system}\textbf{Example of a logical spin represented as a single chain.} Figure
	shows a graphical representation of the Hamiltonian in Eq~(\ref{eq:ham_simple_system}).
	Spins inside the red-dotted box represent the same logical spin/chain while spins outside the box
	belong to other logical spins/chains. The two blue-dashed vertical line are the boundary
	of a cluster growth inside the chain.
	}
\end{figure}

To overcome this limitation, we devised a variant
of the Wolff cluster method for the embedded SK Hamiltonian
which takes into account the existence of the logical spins as a chain of physical spins,
but it does not interfere with the equilibration of the underlying SKM. In our variant, clusters in the
embedded SK Hamiltonian are created by using the following rules:

\begin{enumerate}
	\item Chose a random spin $i$ which represents the ``center'' of the cluster $C$.
	\item Neighbors of the center which belong to the \emph{same logical spin/chain}
			are included as members of the cluster with a probability
			$p(\sigma_i,\,\sigma_j) = 1 - \exp(-\beta|J_F|(1+\sigma_i\sigma_j))$
			(recall that all the intra-chain couplings are always ferromagnetic and
			of magnitude $J_F$).
			New neighbors are considered if that particular pair were not considered before.
	\item The creation process continues until no new neighbors are added. Since $C$ can grow only
			\emph{inside} the logical spin/chain, any cluster can be seen as a connected sub-chain which contains
			the center of the cluster $i$.
	\item Flip the whole cluster with a probability
		$$
			\tilde{p}(\sigma,\,C) =
				\min\left\lbrace 1,\,
					\exp\left(2\beta\!\!\!\!\!\!
					\sum_{\left\langle i,\,j\right\rangle\in\partial^\prime C}
					\!\!\!\!\!J_{ij}\sigma_i\sigma_j
					\right)\!\!
				\right\rbrace,
		$$
		where $\sigma$ is the spin configuration of the system and
		$\partial^\prime C$ consists in all the couplings between spins in $C$ and spins
		which does \emph{not} belong to the same logical spin/chain of the spins in $C$.
\end{enumerate}
Following the same analysis used for the Wolff method \cite{wang1990cluster},
it is straightforward to show
that the above procedure still satisfies the detailed balance: indeed, the limitation that
a cluster can grow only inside a logical spin/chain is balanced by adding the probability
$\tilde{p}(\sigma,\,C)$ to flip the cluster, which involves only extra-chain couplings.
To make this point more clear, consider the simple system depicted in Fig.~(\ref{fig:simple_system})
and
described by the Hamiltonian
\begin{equation}\label{eq:ham_simple_system}
	H=-|J_F|\sum_{i}\sigma_i\sigma_{i+1} + \sum_{(i,\,\alpha)} J_{i\alpha}\sigma_i\sigma_\alpha,
\end{equation}
where $\sigma_i$ represent spins in the same logical spin/chain while $\sigma_\alpha$
represent spins in other logical spins/chains.
Since clusters can be only created within the chain, clusters can be seen as sub-chain
as depicted in Fig.~(\ref{fig:simple_system}). Therefore, the probability to create a certain cluster $C$
can be easily computed and results to be
\begin{subequations}\begin{align}
	p_\text{cluster}(\sigma_i,\,C) &= \nonumber\\
		&\prod_{a\leq i<b}\left(
			1 - e^{-\beta(1 + \sigma_i\sigma_{i+1})}
		\right)\cdot\label{eq:p_simple_system1}\\
		&\cdot e^{-\beta|J_F|\left(2+\sigma_{a-1}\sigma_a+\sigma_b\sigma_{b+1}\right)},
\end{align}\end{subequations}
where $a$ and $b$ are the ends of the sub-chain.
It is important to observe that $p_\text{cluster}(\sigma_i,\,C)$
depends only on the intra chain spins and that
the term in Eq.~(\ref{eq:p_simple_system1}) is invariant by flipping the whole cluster $C$.
The transition probability to flip the cluster $C$
using our modified cluster algorithm is
\begin{align}
	W(\sigma\to\sigma^\prime,\,C) &= p_\text{cluster}(\sigma_i,\,C)\,\tilde{p}(\sigma,\,C),
\end{align}
where $\sigma$ and $\sigma^\prime$ are respectively the spin configurations of the system before and after
to flip the cluster $C$. Therefore, the detailed balance is satisfied if
\begin{align}\label{eq:det_balance_simple_system}
		\frac{p_\text{cluster}(\sigma_i,\,C)}{p_\text{cluster}(\sigma^\prime_i,\,C)}\cdot
		\frac{\tilde{p}(\sigma,\,C)}{\tilde{p}(\sigma^\prime,\,C)} =
		e^{-\beta \Delta H},
\end{align}
with $\Delta H = H(\sigma^\prime) - H(\sigma)$ and $H(\sigma)$ as in Eq.~(\ref{eq:ham_simple_system}).
Since only spins inside the cluster $C$ are flipped, after some calculation one finds that
\begin{subequations}\begin{align}
	\frac{p_\text{cluster}(\sigma_i,\,C)}{p_\text{cluster}(\sigma^\prime_i,\,C)} &=
		e^{-2\beta |J_F|\left(\sigma_{a-1}\sigma_a + \sigma_b\sigma_{b+1}\right)}\\
	\frac{\tilde{p}(\sigma,\,C)}{\tilde{p}(\sigma^\prime,\,C)} &=
	e^{2\beta\sum_{(a\leq i\leq b,\,\alpha)}J_{i\alpha}\sigma_i\sigma_\alpha},
\end{align}\end{subequations}
and therefore, the detailed balance in Eq.~(\ref{eq:det_balance_simple_system})
is satisfied.

In the right panel of Fig.~(\ref{fig:equilib_SK}) we show the convergence of the Binder ratio $g$
for the embedded SK model with $N = 24$ logical spins and $J_F = \sqrt{24}$,
by using our variant of the Wolff cluster method.
In our cluster method, a sweep is defined as a complete update of the system where
all the spins have been chosen as center of a cluster.
Correctly, for a sufficiently large number of sweeps, the Binder ratio $g$ computed
by using our cluster method converges to the same Binder ratio $g$ of the logical SK
model.

Finally, in order to be sure that our cluster method has reached the full equilibration for the
calculation of the Binder ratio $g$, we used the same test as originally proposed in
\cite{bhatt1985search}. In particular, we compute both a lower-bound estimation
$\chi_\text{SG} = N_{\text{hw}}\left\langle q^2 \right\rangle$, where $N_{\text{hw}}$ is the number
of the hardware spins (i.e. $N^2/4+N$), and an upper-bound estimation
$\chi^\prime_\text{SG} = \frac{1}{N_{\text{hw}}}\left\langle
\left[
\sigma_i^{(1)}(\mathcal{T}_{\text{ann}}+t_0)\,\sigma_i^{(1)}(t_0)
\right]
\right\rangle$, with $t_0$ and $\mathcal{T}_{\text{ann}}$ respectively the
equilibration and the measurement time, of the true spin glass susceptibility.
Results are shown in Fig.~(\ref{fig:diff_binder}). Since the two curves
do not coincide for $T < 0.2$ (shaded regions in the figures),
results on the Binder ratio are reliable only for rescaled temperatures $T > 0.2$.

\subsection{\normalsize{Comparison of Embedding of the Sherrington-Kirkpatrick Model with Edwards-Anderson 2D model}}

\begin{figure*}[t]
\centering
\includegraphics[width=0.9\textwidth]{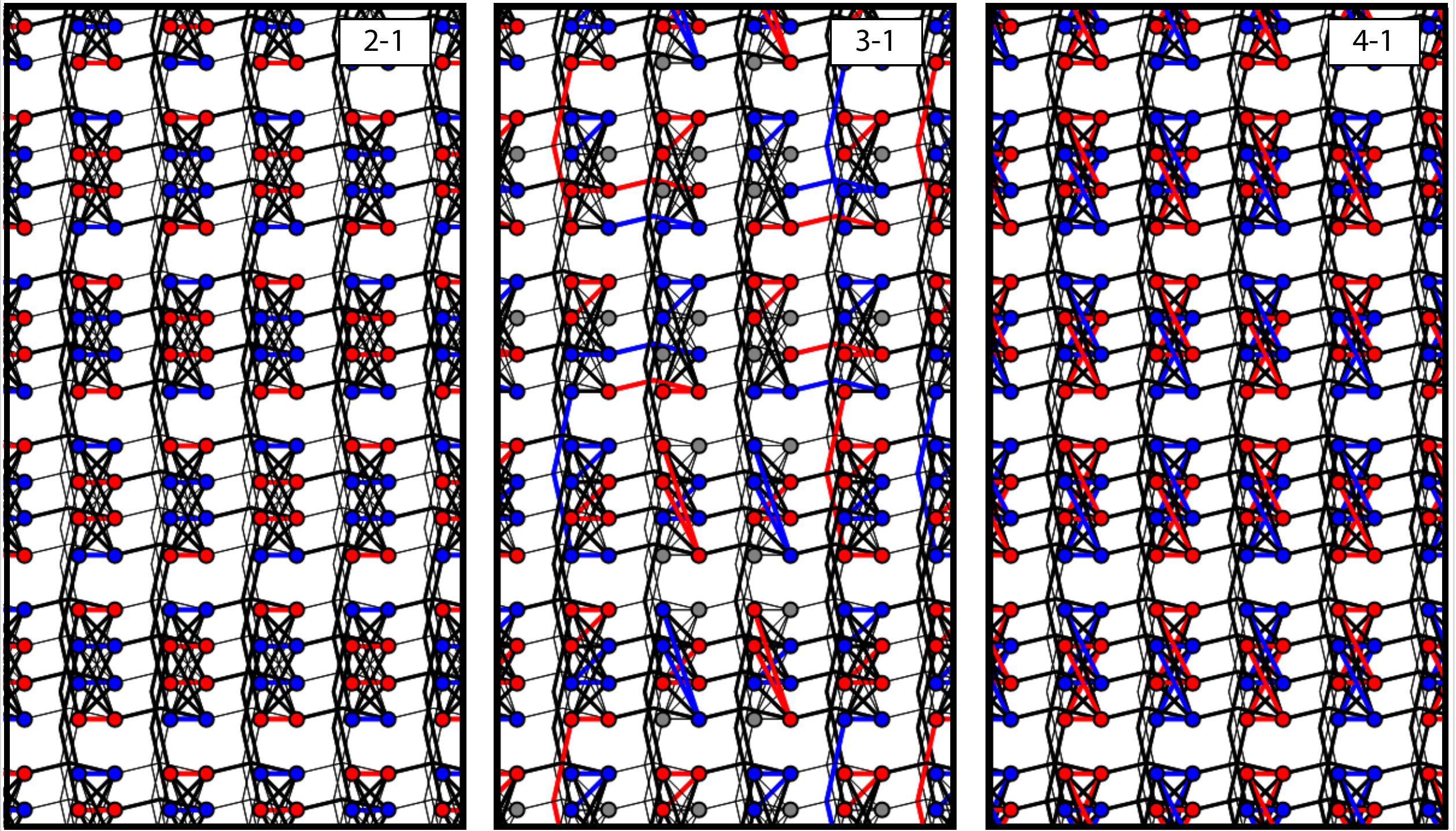}
	\caption{\label{fig:embedding2D}\textbf{Different embeddings for the Edwards-Anderson 2D model.}
		Left: 2-qubits represent one LB, Center: 3-qubits represent one LB, Right: 4-qubits represent one LB. Colored links indicate ferromagnetic couplings.}
\end{figure*}

\begin{figure}[t!]
\includegraphics[width=0.5\textwidth]{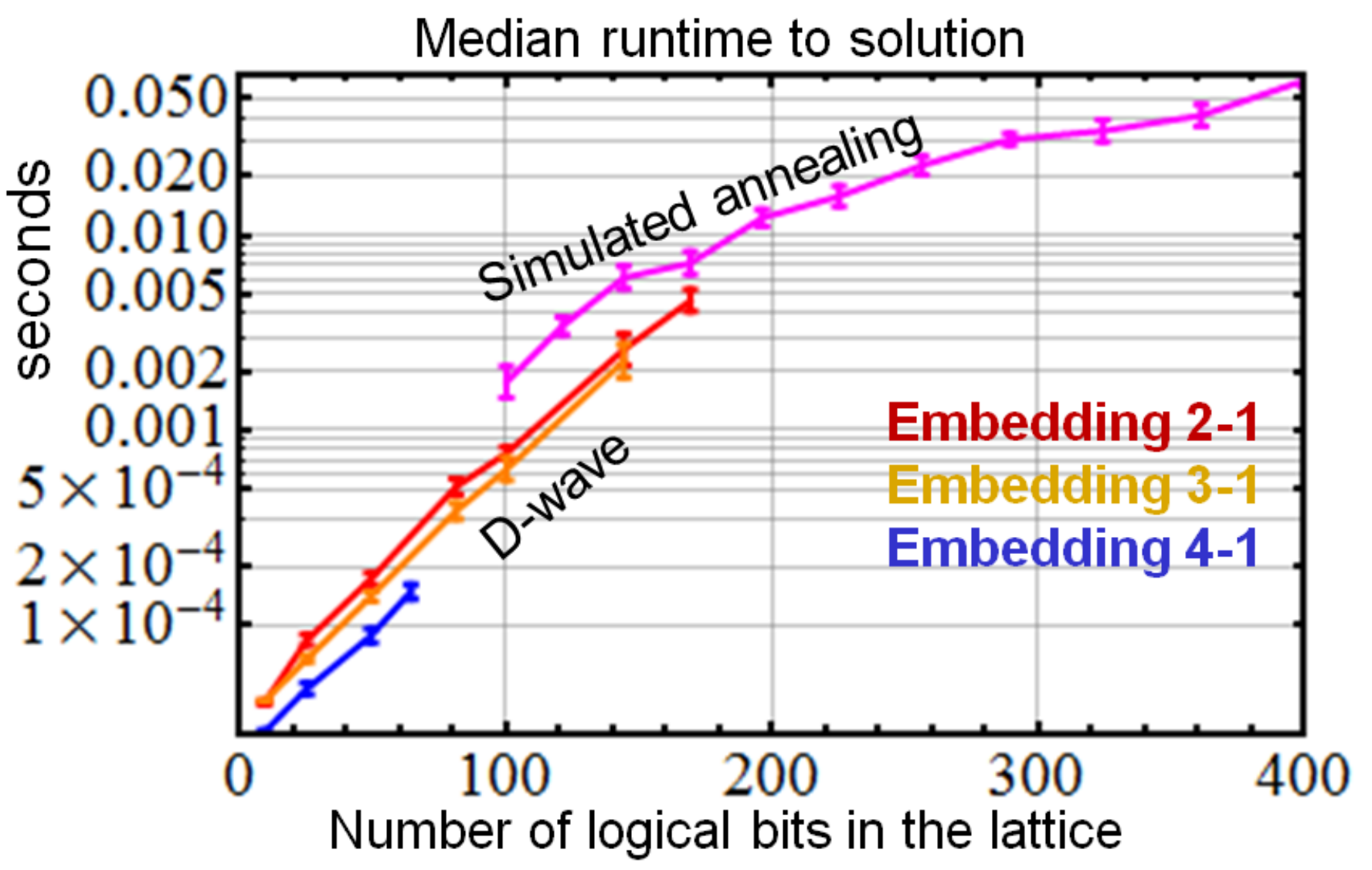}
	\caption{\label{fig:scaling2D}\textbf{Scaling of performance for Edwards-Anderson 2D model.}
Median runtime in seconds between different embeddings, compared with SA. Run protocols and expected runtimes
have been calculated as in the main text for the SKM.}
\end{figure}

Many of the properties which are discussed in the main text concerning optimal
parameter setting are relevant only for embeddings whose average component
size increases with the system size. In this section we program on the D-Wave device
the Edwards Anderson 2D model (2D-EAM), which consists of a disordered Ising model on a square lattice without local fields:

\begin{align}\label{eq:ea}
H^{2D}_{EA}=\sum_{<i,j>} J_{ij}S_{i} S_{j}
\end{align}
We follow the same steps as in the main text to shed light on the differences.
The embedding is straightforward as shown in Fig.~\ref{fig:embedding2D}, where we chose to encode each LB in
ferromagnetic chains of $2,\,3$ or $4$ qubits. The embedding overhead this time is $linear$ instead of quadratic
(as in the SKM) because of the finite connectivity of the lattice to be embedded.
The optimal $J_{F}$ for a given embedding does not scale with the system size, and it is $J_{F}=1.25,\,1.6$ and $2.5$
respectively for the embeddings with $2,\,3$ and $4$ qubits. These values do not match the optimal values found in Fig.~2
of the main paper for the same embedding chain length (i.e. yellow, green and gray lines of Fig.~2 of the main paper),
in accordance with the conjecture that the optimal $J_{F}$ depends on the interplay between the LB criticality and
the problem criticality, which for disordered 2D Ising models doesn't scale with the system size~\cite{pich1998critical}
(however note that
for these small sizes of LB chains we are dominated by finite-size effects so criticality is very loosely defined).

Unsurprisingly, the effect of the error-correction is not as dramatic as in the SKM, but it becomes more pronounced
as the number of qubits in each LB increases.
The scaling of performance (see Fig.~\ref{fig:scaling2D}) does not seem affected by the embedding choice although
larger size ought to be considered in order to make definite statements (which is not possible in current machines).
Comparison with SA (optimized in the linear schedule similarly as we did for the SKM. We checked that the starting
temperature T=1 was close to an optimal choice) is included for sake of completeness. The largest number of logical
spins we considered is $N = 400$.

\vfill

\bibliographystyle{unsrt}
\bibliography{bibliomandra,embedding_SK_suppInfo}{}

\begin{thebibliography}{10}

\bibitem{lucas2014ising}
Andrew Lucas.
\newblock Ising formulations of many np problems.
\newblock {\em Frontiers in Physics}, 2:5, 2014.

\bibitem{smelyanskiy2012near}
Vadim~N Smelyanskiy, Eleanor~G Rieffel, Sergey~I Knysh, Colin~P Williams,
  Mark~W Johnson, Murray~C Thom, William~G Macready, and Kristen~L Pudenz.
\newblock A near-term quantum computing approach for hard computational
  problems in space exploration.
\newblock {\em arXiv preprint arXiv:1204.2821}, 2012.

\bibitem{fu1986application}
Yaotian Fu and Philip~W Anderson.
\newblock Application of statistical mechanics to np-complete problems in
  combinatorial optimisation.
\newblock {\em Journal of Physics A: Mathematical and General}, 19(9):1605,
  1986.

\bibitem{santoro2002theory}
Giuseppe~E Santoro, Roman Marto{\v{n}}{\'a}k, Erio Tosatti, and Roberto Car.
\newblock Theory of quantum annealing of an ising spin glass.
\newblock {\em Science}, 295(5564):2427--2430, 2002.

\bibitem{morita2008mathematical}
Satoshi Morita and Hidetoshi Nishimori.
\newblock Mathematical foundation of quantum annealing.
\newblock {\em Journal of Mathematical Physics}, 49(12):125210, 2008.

\bibitem{brooke1999quantum}
J~Brooke, D~Bitko, G~Aeppli, et~al.
\newblock Quantum annealing of a disordered magnet.
\newblock {\em Science}, 284(5415):779--781, 1999.

\bibitem{zagoskin2014test}
Alexandre~M Zagoskin, Evgeni Ilichev, Miroslav Grajcar, Joseph~J Betouras, and
  Franco Nori.
\newblock How to test the" quantumness" of a quantum computer?
\newblock {\em arXiv preprint arXiv:1401.2870}, 2014.

\bibitem{bunyk2014architectural}
PI~Bunyk, E~Hoskinson, MW~Johnson, E~Tolkacheva, F~Altomare, AJ~Berkley,
  R~Harris, JP~Hilton, T~Lanting, and J~Whittaker.
\newblock Architectural considerations in the design of a superconducting
  quantum annealing processor.
\newblock {\em arXiv preprint arXiv:1401.5504}, 2014.

\bibitem{mcgeoch2013experimental}
Catherine~C McGeoch and Cong Wang.
\newblock Experimental evaluation of an adiabiatic quantum system for
  combinatorial optimization.
\newblock In {\em Proceedings of the ACM International Conference on Computing
  Frontiers}, page~23. ACM, 2013.

\bibitem{DashCPLEX}
S.~Dash.
\newblock A note on qubo instances defined on chimera graphs.
\newblock {\em http://arxiv.org/pdf/1306.1202v2.pdf}, 2013.

\bibitem{boixo2013quantum}
Sergio Boixo, Troels~F R{\o}nnow, Sergei~V Isakov, Zhihui Wang, David Wecker,
  Daniel~A Lidar, John~M Martinis, and Matthias Troyer.
\newblock Quantum annealing with more than one hundred qubits.
\newblock {\em arXiv preprint arXiv:1304.4595}, 2013.

\bibitem{vinci2014distinguishing}
Walter Vinci, Tameem Albash, Anurag Mishra, Paul~A Warburton, and Daniel~A
  Lidar.
\newblock Distinguishing classical and quantum models for the d-wave device.
\newblock {\em arXiv preprint arXiv:1403.4228}, 2014.

\bibitem{ronnow2014defining}
Troels~F. Rønnow, Zhihui Wang, Joshua Job, Sergio Boixo, Sergei~V. Isakov,
  David Wecker, John~M. Martinis, Daniel~A. Lidar, and Matthias Troyer.
\newblock Defining and detecting quantum speedup.
\newblock {\em Science}, 2014.

\bibitem{smolin2013classical}
John~A Smolin and Graeme Smith.
\newblock Classical signature of quantum annealing.
\newblock {\em arXiv preprint arXiv:1305.4904}, 2013.

\bibitem{APO2014QADMF}
Alejandro {Perdomo-Ortiz}, Joseph Fluegemann, Sriram Narasimhan, Vadim
  Smelyanskiy, and Rupak Biswas.
\newblock A quantum annealing approach for fault detection and diagnosis of
  graph-based systems.
\newblock {\em To be submitted}, 2014.

\bibitem{planningquantum}
Eleanor Rieffel, Davide Venturelli, Bryan O'Gorman, Minh Do, Elicia Prystay,
  and Vadim Smelyanskiy.
\newblock A case study in programming a quantum annealer for hard operational
  planning problems.
\newblock {\em To be submitted}, 2014.

\bibitem{perdomo2012finding}
Alejandro Perdomo-Ortiz, Neil Dickson, Marshall Drew-Brook, Geordie Rose, and
  Al{\'a}n Aspuru-Guzik.
\newblock Finding low-energy conformations of lattice protein models by quantum
  annealing.
\newblock {\em Scientific reports}, 2, 2012.

\bibitem{bian2012experimental}
Zhengbing Bian, Fabian Chudak, William~G Macready, Lane Clark, and Frank
  Gaitan.
\newblock Experimental determination of ramsey numbers.
\newblock {\em arXiv preprint arXiv:1201.1842}, 2012.

\bibitem{ogorman2014baysnet}
Bryan O'Gorman, Ryan Babbush, Alejandro Perdomo-Ortiz, Alan Aspuru-Guzik, and
  Vadim Smelyanskiy.
\newblock Bayesian network structure learning using quantum annealing.

\bibitem{PhysRevLett.35.1792}
David Sherrington and Scott Kirkpatrick.
\newblock Solvable model of a spin-glass.
\newblock {\em Phys. Rev. Lett.}, 35:1792--1796, Dec 1975.

\bibitem{APO2014DWtuning}
Alejandro {Perdomo-Ortiz}, Joseph Fluegemann, Vadim~N. Smelyanskiy, and Rupak
  Biswas.
\newblock Programming and solving real-world applications on a quantum
  annealing device.
\newblock {\em To be submitted}, 2014.

\bibitem{harris2010experimental}
Richard Harris, MW~Johnson, T~Lanting, AJ~Berkley, J~Johansson, P~Bunyk,
  E~Tolkacheva, E~Ladizinsky, N~Ladizinsky, T~Oh, et~al.
\newblock Experimental investigation of an eight-qubit unit cell in a
  superconducting optimization processor.
\newblock {\em Physical Review B}, 82(2):024511, 2010.

\bibitem{johnson2010scalable}
MW~Johnson, P~Bunyk, F~Maibaum, E~Tolkacheva, AJ~Berkley, EM~Chapple, R~Harris,
  J~Johansson, T~Lanting, I~Perminov, et~al.
\newblock A scalable control system for a superconducting adiabatic quantum
  optimization processor.
\newblock {\em Superconductor Science and Technology}, 23(6):065004, 2010.

\bibitem{kaminsky2004scalable}
William~M Kaminsky and Seth Lloyd.
\newblock Scalable architecture for adiabatic quantum computing of np-hard
  problems.
\newblock In {\em Quantum computing and quantum bits in mesoscopic systems},
  pages 229--236. Springer, 2004.

\bibitem{eppstein2009finding}
David Eppstein.
\newblock Finding large clique minors is hard.
\newblock {\em J. Graph Algorithms Appl.}, 13(2):197--204, 2009.

\bibitem{Cai-14}
Jun Cai, Bill Macready, and Aidan Roy.
\newblock A practical heuristic for finding graph minors.
\newblock {\url arXiv:quant-ph/1406:2741}, 2014.

\bibitem{choi2011minor}
Vicky Choi.
\newblock Minor-embedding in adiabatic quantum computation: Ii. minor-universal
  graph design.
\newblock {\em Quantum Information Processing}, 10(3):343--353, 2011.

\bibitem{klymko2013adiabatic}
Christine Klymko, Blair~D Sullivan, and Travis~S Humble.
\newblock Adiabatic quantum programming: minor embedding with hard faults.
\newblock {\em Quantum Information Processing}, pages 1--21, 2013.

\bibitem{johnson2011quantum}
MW~Johnson, MHS Amin, S~Gildert, T~Lanting, F~Hamze, N~Dickson, R~Harris,
  AJ~Berkley, J~Johansson, P~Bunyk, et~al.
\newblock Quantum annealing with manufactured spins.
\newblock {\em Nature}, 473(7346):194--198, 2011.

\bibitem{choi2008minor}
Vicky Choi.
\newblock Minor-embedding in adiabatic quantum computation: I. the parameter
  setting problem.
\newblock {\em Quantum Information Processing}, 7(5):193--209, 2008.

\bibitem{error-corrected-lidar}
Kristen~L. Pudenz, Tameem Albash, and Daniel~A. Lidar.
\newblock {Error-corrected quantum annealing with hundreds of qubits}.
\newblock {\em Nature Communications}, 5, February 2014.

\bibitem{young2013adiabatic}
Kevin~C Young, Robin Blume-Kohout, and Daniel~A Lidar.
\newblock Adiabatic quantum optimization with the wrong hamiltonian.
\newblock {\em Physical Review A}, 88(6):062314, 2013.

\bibitem{suzuki2013quantum}
Sei Suzuki, Jun-ichi Inoue, and Bikas~K Chakrabarti.
\newblock {\em Quantum Ising phases and transitions in transverse Ising
  models}.
\newblock 2013.

\bibitem{trevor}
Lanting Trevor.
\newblock private communication.
\newblock {\em D-Wave systems}.

\bibitem{boixo2014evidence}
Sergio Boixo, Troels~F R{\o}nnow, Sergei~V Isakov, Zhihui Wang, David Wecker,
  Daniel~A Lidar, John~M Martinis, and Matthias Troyer.
\newblock Evidence for quantum annealing with more than one hundred qubits.
\newblock {\em Nature Physics}, 10(3):218--224, 2014.

\bibitem{glassychimera}
Helmut~G. Katzgraber, Firas Hamze, and Ruben~S. Andrist.
\newblock Glassy chimeras could be blind to quantum speedup: Designing better
  benchmarks for quantum annealing machines.
\newblock {\em Phys. Rev. X}, 4:021008, Apr 2014.

\bibitem{bhatt1985search}
RN~Bhatt and AP~Young.
\newblock Search for a transition in the three-dimensional j ising spin-glass.
\newblock {\em Physical review letters}, 54(9):924, 1985.

\bibitem{kirkpatrick1983optimization}
Scott Kirkpatrick, MP~Vecchi, et~al.
\newblock Optimization by simmulated annealing.
\newblock {\em Science}, 220(4598):671--680, 1983.

\bibitem{thouless1977solution}
DJ~Thouless, PW~Anderson, and RG~Palmer.
\newblock Solution of'solvable model of a spin glass'.
\newblock {\em Philosophical Magazine}, 35(3):593--601, 1977.

\bibitem{edwards1975theory}
Samuel~Frederick Edwards and Phil~W Anderson.
\newblock Theory of spin glasses.
\newblock {\em Journal of Physics F: Metal Physics}, 5(5):965, 1975.

\bibitem{sherrington1975solvable}
David Sherrington and Scott Kirkpatrick.
\newblock Solvable model of a spin-glass.
\newblock {\em Physical review letters}, 35(26):1792, 1975.

\bibitem{fischer1975static}
KH~Fischer.
\newblock Static properties of spin glasses.
\newblock {\em Physical Review Letters}, 34(23):1438, 1975.

\bibitem{parisi1987spin}
Giorgi Parisi, M~Mezard, and MA~Virasoro.
\newblock Spin glass theory and beyond.
\newblock {\em World Scientific}, 1987.

\bibitem{binder1981critical}
K~Binder.
\newblock Critical properties from monte carlo coarse graining and
  renormalization.
\newblock {\em Physical Review Letters}, 47(9):693, 1981.

\bibitem{katzgraber2014glassy}
Helmut~G Katzgraber, Firas Hamze, and Ruben~S Andrist.
\newblock Glassy chimeras could be blind to quantum speedup: Designing better
  benchmarks for quantum annealing machines.
\newblock {\em Physical Review X}, 4(2):021008, 2014.

\bibitem{dutta2011transverse}
Amit Dutta, Uma Divakaran, Diptiman Sen, Bikas~K Chakrabarti, Thomas~F
  Rosenbaum, and Gabriel Aeppli.
\newblock Transverse field spin models: From statistical physics to quantum
  information.
\newblock {\em ChemInform}, 42(18), 2011.

\bibitem{parisi1993several}
G~Parisi, F~Ritort, and F~Slanina.
\newblock Several results on the finite-size corrections in the
  sherrington-kirkpatrick spin-glass model.
\newblock {\em Journal of Physics A: Mathematical and General}, 26(15):3775,
  1993.

\bibitem{aspelmeier2008finite}
T~Aspelmeier, A~Billoire, E~Marinari, and MA~Moore.
\newblock Finite-size corrections in the sherrington--kirkpatrick model.
\newblock {\em Journal of Physics A: Mathematical and Theoretical},
  41(32):324008, 2008.

\bibitem{billoire2011finite}
A~Billoire, LA~Fernandez, A~Maiorano, E~Marinari, V~Martin-Mayor, and
  D~Yllanes.
\newblock Finite-size scaling analysis of the distributions of pseudo-critical
  temperatures in spin glasses.
\newblock {\em Journal of Statistical Mechanics: Theory and Experiment},
  2011(10):P10019, 2011.

\bibitem{wolff1989collective}
Ulli Wolff.
\newblock Collective monte carlo updating for spin systems.
\newblock {\em Physical Review Letters}, 62(4):361, 1989.

\bibitem{wang1990cluster}
Jian-Sheng Wang and Robert~H Swendsen.
\newblock Cluster monte carlo algorithms.
\newblock {\em Physica A: Statistical Mechanics and Its Applications},
  167(3):565--579, 1990.

\bibitem{pich1998critical}
C~Pich, AP~Young, H~Rieger, and N~Kawashima.
\newblock Critical behavior and griffiths-mccoy singularities in the
  two-dimensional random quantum ising ferromagnet.
\newblock {\em Physical review letters}, 81(26):5916, 1998.

\end{thebibliography}

\end{document}